\documentclass[aps,prd,showpacs, showkeys, preprint, preprintnumbers, amsmath, amssymb,superscriptaddress,nofootinbib,eqsecnum]{revtex4-2}
\usepackage{graphicx}
\usepackage{subcaption}
\usepackage{txfonts}
\usepackage{dcolumn}
\usepackage{mathrsfs}
\usepackage{bm}
\usepackage[title]{appendix}
\usepackage{amsmath, amssymb, epsfig, float, diagbox}
\usepackage{color}
\allowdisplaybreaks[3]
\usepackage{hyperref}
\usepackage{graphicx}
\usepackage{orcidlink}
\newcommand{\bea}{\begin{eqnarray*}}
\newcommand{\eea}{\end{eqnarray*}}
\newcommand{\bean}{\begin{eqnarray}}
\newcommand{\eean}{\end{eqnarray}}

\begin{document}

\title{Energy flux and waveform of gravitational wave generated by coalescing slow-spinning binary system in effective one-body theory}

\author{Weike Deng}
%\email[ ]{1721458793@qq.com}
\affiliation{Department
	of Physics, Key Laboratory of Low Dimensional Quantum Structures and
	Quantum Control of Ministry of Education, and Synergetic Innovation
	Center for Quantum Effects and Applications, Hunan Normal
	University, Changsha, Hunan 410081, P. R. China}

\author{Sheng Long\,\orcidlink{0009-0009-0163-2724}
	\footnote{Sheng Long and Weike Deng contributed equally to this work and should be considered as co-first author.}}
%\email[ ]{643351248@qq.com}
\affiliation{Department
	of Physics, Key Laboratory of Low Dimensional Quantum Structures and
	Quantum Control of Ministry of Education, and Synergetic Innovation
	Center for Quantum Effects and Applications, Hunan Normal
	University, Changsha, Hunan 410081, P. R. China}

\author{Jiliang {Jing} \,\orcidlink{0000-0002-2803-7900}
	\footnote{Corresponding author, jljing@hunnu.edu.cn}}
\affiliation{Department of Physics, Key Laboratory of Low Dimensional Quantum Structures and Quantum Control of Ministry of Education, and Synergetic Innovation
	Center for Quantum Effects and Applications, Hunan Normal
	University, Changsha, Hunan 410081, P. R. China}
\affiliation{Center for Gravitation and Cosmology, College of Physical Science and Technology, Yangzhou University, Yangzhou 225009, P. R. China}

\begin{abstract}
 
We extend our research on the energy flux and waveform characteristics of gravitational waves generated by merging nonspinning binary black holes through self-consistent effective one-body theory \cite{L2023} to include binary systems with slowly spinning black holes. Initially, we decompose the equation for the null tetrad component of the gravitationally perturbed Weyl tensor $\psi^B_{4}$ into radial and angular parts, leveraging the second-order approximation of the rotation parameter $a$. Subsequently, we derive an analytical solution for the radial equation and observe that our results are contingent upon the parameters $a_2$, $a_3$ and $a$, which represent the second- and third-order correction parameters, respectively. Ultimately, we calculate the energy flux, the radiation--reaction force and the waveform for the ``plus" and ``cross" modes of the gravitational waves generated by merging slowly spinning binary black holes.

\end{abstract}


\maketitle

\newpage

	\section{Introduction}\label{section1}
The concept of gravitational waves (GWs) can be traced back to general relativity~\cite{Einstein1916}. However, due to a lack of sufficient experimental observations for verification, research in this field did not attract much attention in the decades following Einstein's prediction. Interest only surged into the 1960s, following Bondi and Sachs' established formalism, which provided the theoretical foundation for the existence of GWs. The amplitude of GWs is closely related to mass, positioning black holes as a primary source. The metric for a spherically symmetric non-symmetric black hole was first derived by Schwarzschild and later generalized to include rotation by Kerr. In 1963, Schiffer~\cite{Schiffer1963} presented a specific derivation of both geometries by simplifying the Einstein free space field equations for the algebraically special form of the Kerr metric.

The detection of GWs began with Weber's resonators in the 1960s, employing techniques such as environmental perturbation exclusion and signal amplification, which have propelled advancements in GW detection. The first direct detection of a GW signal, GW150914~\cite{Abbott2016}, from the merger of two black holes, was reported by the Laser Interferometer Gravitational Wave Observatory (LIGO) Scientific Collaboration and the Virgo Collaboration. Subsequently, the Advanced LIGO (aLIGO)~\cite{Abbott2015}, later joined by the Advanced Virgo~\cite{Acernese2015} in 2017, has detected numerous coalescing binary black hole events~\cite{Abbott6,Abbott7,Abbott8,Abbott9,Abbott10,Abbott11,Abbott12,Abbott13,Abbott14} and two binary neutron star events ~\cite{Abbott15,Abbott16} from its $O1$ to $O3$ observation runs. Among these events, coalescing compact object binary systems are undoubtedly the most promising sources. Therefore, a rigorous examination of all aspects of black-hole research is imperative~\cite{ Zhoux, Liux, Chens, Chens1, Zhangx, Liuw, Jing2021}. The successful detection of these GW signals marks the arrival of a new generation of GW astronomy, opening unprecedented opportunities for humans to explore cosmology, especially concerning extreme physical processes and phenomena, such as strong gravitational fields, extremely dense celestial bodies, and high-energy processes.

Despite GWs being significantly weaker than background noise, the GWs emitted by compact binary systems carry crucial information about their sources. This information can be extracted from the noise using the matched filtering technique, which may require tens of thousands of waveform templates. However, the high computational cost and the large parameter space associated with spinning binary black holes make it impractical to rely solely on numerical relativity for generating a comprehensive template bank of gravitational waveforms~\cite{Pan2010}. Therefore, the effective-one-body (EOB)~\cite{Buo1999,Dam2000,Dam2009,Buo2000,Dam2001} theory was introduced in 1999 by Buonanno and Damour. Based on the post-Newtonian approximation, as a novel approach to investigate the general gravitational radiation generated by coalescing compact object binary systems. The synergy between the EOB theory and numerical relativity has played an essential role in the data analysis of GWs. Building on this great success, Damour~\cite{Dam2016} developed another EOB theory with a post-Minkowskian (PM) approximation in 2016. Unlike its predecessor, this new approach removes the restriction that $ v/c$ must be small, leading to a significant expansion in related research~\cite{Bini2017,Bini2018,Dam2018,Ant2019,Dam2020,Bini2020,He2016,He2019,Blan2018,Che2018,Vine2019,Cris2019,Coll2019,Plef2019,Bini20201,Che2020,Bern2019,Bern20191}. In this context, we have developed a self-consistent effective-one-body (SCEOB) theory~\cite{Jing2022} tailored for the real spinless two-body system under a PM approximation. This theory is specifically designed to explore the dynamics and GW emissions of merging nonspinning black holes characterized by two mass parameters ($m_1,m_2$). We are now looking to extend the application of the SCEOB theory to binary systems of spinning black holes, taking into account both mass and spin parameters ($m_1,S_1,m_2,S_2$).

In previous work~\cite{Jing2023}, following the approach of Damour and Barausse et al.~\cite{Bar2011,Dam20011}, we successfully developed an effective rotating metric for real spin two-body systems and constructed an improved SCEOB Hamiltonian for scenarios where a spinning test particle orbits a massive rotating black hole, as defined by the specified metric.
Key to detecting the "plus" and "cross" modes of GWs is the calculation of the radiation--reaction force (RRF), which requires analysis of the GW energy flux. This analysis hinges on solving for the null tetrad component of the gravitationally perturbed Weyl tensor $\psi^B_{4}$ since it is related to the two modes of the GWs expressed as $\psi_{4}^{B}=\frac{1}{2}(\ddot{h}_{+}-i\ddot{h}_{\times})$ at infinity. However, the challenge lies in the decoupled equation within this effective spacetime that cannot be separated naturally. Yet, insights from the LIGO--Virgo Consortium indicate that most events detected during the $O1$ and $O2$ observation involve black holes with low effective spin values. Furthering this, Roulet and Zaldarriaga~\cite{Rou2019}, through a comprehensive reanalysis of LIGO--Virgo strain data and models for angular spin distributions, found that the spin systems predominantly consist of slowly spinning black holes, namely $a<0.1$. A notable example is GW190814, involving a merger between a $23M_{\odot}$ black hole and a $2.6M_{\odot}$ compact object, which enforced tight constraints on the spin of the primary black hole to $a<0.07$ owing to the large mass ratio. These discussions enable us to divide the decoupled equation into the radial and angular parts in scenarios involving slow rotations~\cite{Koj1992, Pao2012,Sim2020,Jim2019,Chris2021}. Building on the initiatives of Damour--Nagar--Pan~\cite{Pan2010,Buo20091,Dam200911,Dam2007}, we also derived a formal solution for $\psi^B_4$ up to the first order of the rotation parameter $a$ using the Green function. This allowed us to delineate the waveform for the plus and cross modes of the GW in the effective rotating spacetime~\cite{Jing2023}.

In this paper, we expand the decoupled equation of the null tetrad component of the perturbed Weyl tensor $\psi^B_{4}$ to include second-order effects in the rotation parameter $a$. By using the definition and characteristics of spin-weighted spherical harmonics, we demonstrate that the decoupled equation can be variably separated into radial and angular parts. Specifically, when the quantum numbers $\ell$ and $m$ are fixed, the angular part related to $\theta$ can be completely transformed into specific numerical values through integral formulas. Next, we rewrite the radial function $R_{\ell m \omega}$ in the form of a Teukolsky-like equation and convert it to a simpler Sasaki--Nakamura-like (S--N) equation. Using the method proposed by Sasaki~\cite{Sa1994} et al., we obtain the corresponding solutions to the S--N-like equation. It is imperative to recognize that our results are contingent upon the parameters $a_2$, $a_3$ and $a$, where the first two correspond to second- and third-order correction parameters, respectively. As $a$ approaches zero, our results align with those from the nonspinning case~\cite{Long:2023vph,L2023}. Setting all three parameters to zero yields results consistent with the simplest Schwarzschild background.

The rest of the paper is organized as follows. In Section II, we review the effective metric, expand the decoupled equation to the second order based on prior work, and separate $\psi^{B}_{4}$ into its radial and angular parts. In Section III, we transform the radial function $R_{\ell m \omega}$ into the S--N-like equation and present its corresponding analytical solution. Section IV applies similar techniques to the tetrad components of the energy--momentum tensor and provides the ultimate expression of $\psi_4^{B}$. In addition, it offers analytical expressions for the reduced RRF and waveform. The final section presents conclusions and discussions.

\section{Variables separation equation of \texorpdfstring{ $\psi^{B}_{4}$}{} in scenarios involving slow rotations}\label{sec:2}
Based on the work of~\cite{Jing2022} and using the approach of constructing an effective rotating metric as presented by Damour and Barausse et al.~\cite{Bar2011,Dam20011}, we have obtained the effective rotating metric for a real spin two-body system that can be described by \cite{Jing2023}
\begin{align}\label{metric}
	d s^2=\frac{\Delta-a^2 \sin ^2 \theta}{\Sigma} d t^2-\frac{\Sigma}{\Delta} d r^2-\Sigma d \theta^2-\frac{\Lambda_t \sin ^2 \theta}{\Sigma} d \phi^2+\frac{2 \omega_j \sin ^2 \theta}{\Sigma} d t d \varphi,
\end{align}
with
\begin{align}
	\nonumber	& \Sigma=\overline{\rho \rho}^*, \quad \bar{\rho}=r+i a \cos \theta, \quad \bar{\rho}^*=r-i a \cos \theta, \quad \Delta=\Delta^0+a^2, \\
	& \Lambda_t=\varpi^4-a^2 \Delta \sin ^2 \theta, \quad \varpi=\left(r^2+a^2\right)^{\frac{1}{2}}, \quad \omega_j=a\left(a^2+r^2-\Delta\right),
\end{align}
where $\Delta^{0}$ is the part that does not contain the rotational parameter $a$, which can be expressed as
\begin{align}
	\Delta^0=r^2-2 G M r+\sum_{i=2}^{\infty} a_i \frac{\left(G M\right)^i}{r^{i-2}},
\end{align}
$M$ is the mass of the black hole, and $a_{i}$ indicates the PM correction parameter whose specific expressions have been calculated up to 4PM order in Ref.~\cite{Jing20231}.

The decoupled equation of $\psi^{B}_{4}$ in the effective metric (\ref{metric}) can be expressed as \cite{Jing2023}
\begin{align}\label{deceq}
	\left[\Delta\left(\mathscr{D}_{-1}^{\dagger}+\frac{2 \overline{\rho \rho}^*}{\Delta} F_1\right)\left(\mathscr{D}_0-\frac{3}{\bar{\rho}^*}\right)+F_2\left(\mathscr{L}_{-1}-\sqrt{2} \bar{\rho}^* F_3\right)\left(\mathscr{L}_2^{\dagger}-\frac{3 i a \sin \theta}{\bar{\rho}^*}\right)+2 \overline{\rho \rho}^* F_4\right] \phi_4^B=\mathcal{T}_4,
\end{align}
with
\begin{align}
	\nonumber \mathcal{T}_4 & =4 \pi G F_4\left\{\mathscr{L}_{-1}\left[\frac{\bar{\rho}^*}{3 \psi_2-2 \phi_{11}} \mathscr{L}_0\left(\overline{\rho \rho}^{* 2} T_{n n}\right)\right]+\frac{\Delta^2}{2} \mathscr{D}_0^{\dagger}\left[\frac{\bar{\rho}^*}{3 \psi_2+2 \phi_{11}} \mathscr{D}_0^{\dagger}\left(\bar{\rho}^{-1} \bar{\rho}^{* 2} T_{\overline{m m}}\right)\right]\right. \\
	& \left.+\frac{\Delta^2}{\sqrt{2}}\left\{\mathscr{D}_0^{\dagger}\left[\frac{\bar{\rho}^{-2} \bar{\rho}^*}{\Delta\left(3 \psi_2+2 \phi_{11}\right)} \mathscr{L}_{-1}\left(\bar{\rho}^2 \bar{\rho}^{* 2} T_{\overline{m} n}\right)\right]+\mathscr{L}_{-1}\left[\frac{\bar{\rho}^{-2} \bar{\rho}^*}{3 \psi_2-2 \phi_{11}} \mathscr{D}_0^{\dagger}\left(\frac{\bar{\rho}^2 \bar{\rho}^{* 2}}{\Delta} T_{\overline{m} n}\right)\right]\right\}\right\},
\end{align}
where $\phi_4^B=(\bar{\rho}^*)^4\psi_4^{B}$, $T_{n n}=\phi_{22}^B /(4 \pi G)$, $T_{n \overline{m}}=\phi_{21}^B /(4 \pi G)$, and $T_{\overline{m m}}=\phi_{20}^B /(4 \pi G)$. The definitions of coefficients $F_{i}\ (i=1,2,3,4)$, two operators $\mathscr{L}$ and $\mathscr{D}$, component of the traceless Ricci tensor $\phi_{11}$, and component of the Weyl tensor $\psi_{2}$ can be found in~\cite{Jing2023}.

To find a solution of $\psi^{B}_{4}$, we should separate the variables of the above-decoupled equation (\ref{deceq}) and determine the tetrad components of the energy--momentum tensor of the system. However, it seems that direct separation of this equation is not feasible.
Since the bulk of the population of the observed binary black holes merger events involve slowly spinning black holes~\cite{Chris2021}, a conclusion further supported by the remarkably strong spin constraint ($a<0.07$) on the $23M_{\odot}$ primary black hole in GW190814~\cite{Abbott13}, we can separate the decoupled equation in slowly rotation cases~\cite{Koj1992, Pao2012,Sim2020,Jim2019,Chris2021}. Under these conditions, which closely approximate spherical symmetry, $\phi^{B}_{4}$ can be decomposed into Fourier-harmonic components according to
\begin{align}\label{psieq}
	\phi_4^B=\sum_{\ell m} \frac{1}{\sqrt{2 \pi}} \int d \omega\, e^{-i(\omega t-m \varphi)}\, { }_{-2} Y_{\ell m}(\theta)\, R_{\ell m}(r),
\end{align}
where the angular function ${ }_{-2} Y_{\ell m}(\theta)$ is called the spin-weighted spherical harmonic that can be normalized as
\begin{align}
	\int_0^\pi{ }_{-2} \, Y_{\ell m}^*(\theta){ }_{-2}\, Y_{\ell m}(\theta) \, \sin \theta\, d \theta=1.
\end{align}
Therefore, we can expand the decoupled equation ($\ref{deceq}$) to second order with respect to $a$, which is expressed as
\begin{align}\label{seceq}
	&\left[A(\ell)+a\, B-a \,\cos \theta\, C(\ell)+a^2\cos \theta\, D+a^2 \cos ^2 \theta\, E(\ell)+a^2 \cos \theta \sin \theta \,F\,\partial_\theta+a^2\, G\right] \phi_4^B\nonumber \\ &=\mathcal{T}^{(0)}_4+a \mathcal{T}^{(1)}_4+a^2 \mathcal{T}^{(2)}_4,
\end{align}
the definitions of these coefficients are
\begin{align}\label{seceq1}
	\nonumber	A(\ell) & =\Delta^0\left[\mathscr{D}_{-1}^{+0} \mathscr{D}_0^0-\left(\frac{3}{r} \mathscr{D}_{-1}^{+0}+\tilde{F}_1^0 \mathscr{D}_0^0\right)+\frac{3}{r^2}\left(1+r \tilde{F}_1^0\right)\right]+\lambda\,F_2^0+2 r^2 F_4^0, \\
	\nonumber	B & =i m\left[\left(\frac{3}{r}-\tilde{F}_1^0-\frac{2\left(\Delta^0\right)_{r}}{\Delta^0}+\frac{2 i r^2 \omega}{\Delta^0}\right)-2 F_2^0\left(i \omega+\frac{3}{r}\right)\right], \\
	\nonumber	C(\ell) & =i \Delta^0\left[\left(\frac{3}{r^2} \mathscr{D}_{-1}^{+0}+\tilde{F}_1^1 \mathscr{D}_0^0\right)-\frac{3}{r^3}\left(2+r \tilde{F}_1^0+r^2 \tilde{F}_1^1\right)\right]+2 F_2^0\left(\omega-\frac{3 i}{r}\right)-i \lambda\, F_2^1-2 i r^2 F_4^1,\\
	\nonumber D&= m\left[\left(\tilde{F}_1^1-\frac{3}{r^2}\right)+F_2^0\left(\frac{3}{r^2}-\tilde{F}_3^2\right)+2 F_2^1\left(i \omega+\frac{3}{r}\right)\right], \\
	\nonumber	E(\ell)&= \Delta^0\left[\frac{3}{r^3}\left(\mathscr{D}_{-1}^{+0}-\frac{3}{r}-\tilde{F}_1^0-r \tilde{F}_1^1+r^2 \tilde{F}_1^2-\frac{r^3}{3} \tilde{F}_1^2 \mathscr{D}_0^0\right)\right] \\
	\nonumber	& +2 F_2^0\left(\frac{\omega^2}{2}-\frac{3}{r^2}+\tilde{F}_3^2-\frac{3 i}{r} \omega\right)-2 F_2^1\left(i \omega+\frac{3}{r}\right)+\lambda\, F_2^2+2\left(F_4^0+r^2 F_4^2\right), \\
	\nonumber	F &= F_2^0\left(\frac{3}{r^2}+\tilde{F}_3^2\right), \\
	\nonumber	G&= -\tilde{F}_1^0\left(\Delta^0 \mathscr{D}_0^2-\frac{3}{r}+\mathscr{D}_0^0\right)+\left(\Delta^0 \mathscr{D}_0^2-\frac{3}{r}\right) \mathscr{D}_{-1}^{+0}+\left(\Delta^0 \mathscr{D}_{-1}^{+2}+\mathscr{D}_{-1}^{+0}\right) \mathscr{D}_0^0 \\
	& +\Delta^0\left[\left(\mathscr{D}_0^2\right)_{r r}+\frac{m^2}{\left(\Delta^0\right)^2}-\frac{3}{r} \mathscr{D}_{-1}^{+2}+\frac{1}{\Delta^0} \frac{3}{r^2}\right]+2 F_2^0\left(-\frac{\omega^2}{2}+\frac{3}{r^2}+\frac{3 i}{r} \omega\right),
\end{align}
where $\lambda=(\ell-1)(\ell+2)$, the superscripts on all these quantities represent the corresponding order in the series expansion with respect to $a$. The functions $F^{i}_{j}$ $(i=0,1,2$ and $j=1,2,3,4)$ are detailed in Appendix $\ref{ApenA}$.

On the other hand, the trigonometric function appearing in Eq.~($\ref{seceq}$) can be expressed using the spin-weighted spherical harmonics as
\begin{align}\label{seceq2}
\sin\theta=2\sqrt{\frac{2\pi}{3}}{}_{1}Y_{1 0}, \quad \cos \theta=2 \sqrt{\frac{\pi}{3}} {}_{0}Y_{1 0},\quad \cos ^2 \theta=\frac{4}{3} \sqrt{\frac{\pi}{5}} {}_{0}Y_{2 0}+\frac{1}{3}.
\end{align}
Multiplying the Eq.~($\ref{seceq}$) by the complex conjugate ${}_{-2}Y^{*}_{\ell m}$ and taking the integrate over the angles, we obtain an integral involving the multiplication of three spherical harmonics, which can be expressed as \cite{M2008}
\begin{align}\label{seceq3}
	\oint{}_{-2} Y^{*}_{\ell m}\, {}_{-2} Y_{\ell m}\, {}_{s^{\prime}} Y_{\ell^{\prime} m^{\prime}}\, d \Omega=\left[\frac{\left(2 \ell+1\right)^{2}\left(2 \ell^{\prime}+1\right)}{4 \pi}\right]^{1 / 2}\left(\begin{array}{ccc}
		\ell & \ell & \ell^{\prime}\\
		2 & 2 & -s^{\prime}
	\end{array}\right)\left(\begin{array}{ccc}
		\ell & \ell & \ell^{\prime}\\
		m & m & m^{\prime}
	\end{array}\right),
\end{align}
the matrices are the Wigner $3-\ell m$ symbols and it has the following general form 
\begin{align}\label{seceq4}
	\nonumber	& \left(\begin{array}{ccc}
		\ell_1 & \ell_2 & \ell_3 \\
		m_1 & m_2 & m_3
	\end{array}\right)=(-1)^{\ell_1-m_1} \delta_{m_1+m_2,-m_3} \\
	\nonumber	& \quad \times\left[\frac{\left(\ell_1+\ell_2-\ell_3\right) !\left(\ell_1+\ell_3-\ell_2\right) !\left(\ell_2+\ell_3-\ell_1\right) !\left(\ell_3+m_3\right) !\left(\ell_3-m_3\right) !}{\left(\ell_1+\ell_2+\ell_3+1\right) !\left(\ell_1+m_1\right) !\left(\ell_1-m_1\right) !\left(\ell_2+m_2\right) !\left(\ell_2-m_2\right) !}\right]^{1 / 2} \\
	& \quad \times \sum_{k \geq 0} \frac{(-1)^k}{k !}\left[\frac{\left(\ell_2+\ell_3+m_1-k\right) !\left(\ell_1-m_1+k\right) !}{\left(\ell_3-\ell_1+\ell_2-k\right) !\left(\ell_3-m_3-k\right) !\left(\ell_1-\ell_2+m_3+k\right) !}\right] ,
\end{align}
the sum runs over all values of $k$ for which the arguments within the factorials are non-negative. Furthermore, if the particular combination of $[\ell_i m_i]$ results in negative arguments for the factorials outside the sum, then the corresponding coefficient vanishes. By using Eqs.~($\ref{seceq}$)-($\ref{seceq4}$), the equation of the radial part can be written as
\begin{align}\label{sepeq}
\left[A(\ell)+a\, B-a\, \mathcal{B}\, C(\ell)+a^2\mathcal{B}\, D+a^2\mathcal{D}\, E(\ell)+a^2\mathcal{H}\,F\right.\left.+a^2 G\right]R_{\ell m \omega}=T_{\ell m \omega},
\end{align}
with
\begin{align}
	\nonumber &\mathcal{B}=\frac{2m}{\ell \left(\ell+1\right)},\\
	\nonumber &\mathcal{D}=\frac{1}{3}+\frac{2}{3}\frac{\left(\ell+4\right)\left(\ell-3\right)\left(\ell^2+\ell-3m^2\right)}{\ell \left(\ell+1\right)\left(2\ell+3\right)\left(2\ell-1\right)},\\
	\nonumber &\mathcal{H}=2\sqrt{\frac{2\pi}{15}}\oint{}_{-2} Y^{*}_{\ell m} \left(\frac{d {}_{-2} Y_{\ell m}}{d\theta}\right){}_{1}Y_{20} d \Omega,\\
	&T_{\ell m \omega}=\frac{1}{2 \pi} \int_{-\infty}^{+\infty} d t \int d \Omega\left(\mathcal{T}_4^{(0)}+a \mathcal{T}_4^{(1)}+a^{2} \mathcal{T}_4^{(2)}\right) e^{i(\omega t-m \varphi)} \frac{{ }_{-2} Y_{\ell m}^*}{\sqrt{2 \pi}} ,
\end{align}
where $T_{\ell m \omega}$ is the source term that should also be expanded in the series of $a$. The explicit form of $\mathcal{T}_4^{(i)}\ (i=0,1,2)$ is described below.

\section{Analytical solution for radial equation of \texorpdfstring{ $R_{\ell m \omega}$}{} without source}\label{sec:3}
In this section, we should first transform the radial equation ($\ref{sepeq}$) that is devoid of source and features a long-range potential into the short-range potential S--N equation to find the corresponding solution. Then, the radial equation incorporating a source can be obtained through the Green function.

\subsection{General discussion for Teukolsky-like equation}
We find that the radial function $R_{lm\omega}$ shown by Eq.~($\ref{sepeq}$) obeys the Teukolsky-like equation in the form
\begin{align}\label{teueq}
	\left[\frac{\Delta^2}{f(r)} \frac{d}{d r}\left(\frac{f(r)}{\Delta} \frac{d}{d r}\right)-V(r)\right] R_{l m \omega}=T_{\ell m \omega} ,
\end{align}
with 
\begin{align}
	\nonumber f(r)=&-\frac{3 GM}{r^3 F^0_{4} }	+\frac{3 i a\mathcal{B} \,GM \left(r F^1_{4} -3 F^0_{4}\right)}{r^4\left(F^0_{4}\right)^2}\\
	&-\frac{3 a^{2}GM}{2 r^3 F^0_{4}}\left(\left(\frac{3i\mathcal{B}}{r}-\frac{i\mathcal{B} F^1_{4}}{F^0_{4}}\right)^{2}+2\left(-\frac{3\mathcal{D}}{2r^2}-\frac{\mathcal{D} \left(F^1_{4}\right)^2}{2\left(F^0_{4}\right)^2}-\frac{\mathcal{D}F^2_{4}}{F^0_{4}}\right)\right).
\end{align}
The potential $V(r)$ can be calculated through Eq.~($\ref{sepeq}$) as
\begin{align}
	V(r)=-(V^0+a V^1+a^2 V^2)
\end{align}
with
{\small \begin{align}
		\nonumber	V^0=&\frac{r^2 \omega\left(r^2 \omega+2 i \Delta^{0^{\prime}}\right)}{\Delta^0}-i r \omega\left(5-r \tilde{F}_{1}^{0}\right)+\frac{3\left[\Delta^0\left(1+r \tilde{F}_{1}^{0}\right)+r \Delta^{0^{\prime}}\right]}{r^2}+2 r^2 F_{4}^{0}-\lambda\,F_{2}^{0},\\
		\nonumber V^1=&B-\mathcal{B}\left\{\left(r^2 \tilde{F}_{1}^{1}-3\right) \omega-\frac{3 i}{r^3}\left[r \Delta^{0^{\prime}}+\Delta^0\left(2+r \tilde{F}_{1}^{0}+r^2 \tilde{F}_{1}^{1}\right)\right]+2 F_{2}^{0}\left(\omega-\frac{3 i}{r}\right)-2 i r^2 F_{4}^{1}+i\lambda\,F_{2}^{1} \right\},\\
		\nonumber V^2=&\mathcal{B}D+\mathcal{D}\left\{\Delta^0\left[\frac{3}{r^3}\left(\frac{i \omega r^2-\Delta^{0^\prime}}{\Delta^0}-\frac{3}{r}-\tilde{F}_1^0-r \tilde{F}_1^1+r^2 \tilde{F}_1^2+\frac{i\omega r^5}{3\Delta^0} \tilde{F}_1^2 \right)\right]+2 F_2^0\left(\frac{\omega^2}{2}-\frac{3}{r^2}+\tilde{F}_3^2-\frac{3 i}{r} \omega\right)\right.\\
		\nonumber&\left.-2 F_2^1\left(i \omega+\frac{3}{r}\right)+\lambda\, F_2^2+2\left(F_4^0+r^2 F_4^2\right)\right\}+\mathcal{H}F+\frac{m^2+2r^2 \omega^2+2 i \omega \Delta^{0^\prime}}{\Delta^0}-\frac{r^2\omega\left(r^2\omega+2i\Delta^{0^\prime}\right)}{(\Delta^{0})^2}\\
		&+\frac{3}{r^2}\left(1-i\omega r\right)+\tilde{F}_1^{0}\left(i \omega+\frac{3}{r}\right)+2 F_2^0\left(-\frac{\omega^2}{2}+\frac{3}{r^2}+\frac{3 i}{r} \omega\right).
	\end{align}
}
On the other hand, the asymptotic solutions for the homogeneous of Eq.~($\ref{teueq}$) are
\begin{align}\label{RBB}
	& R_{a s y}^{\mathrm{in}} \rightarrow \begin{cases}B_{\ell m \omega}^{\mathrm{trans}}\left(\Delta^0\right)^2 e^{-i \omega r^*}, & \text { for } r \rightarrow r_{+}, \\
		r^3 B_{\ell m \omega}^{\mathrm{ref}} e^{i \omega r^*}+r^{-1} B_{\ell m \omega}^{\mathrm{inc}} e^{-i \omega r^*}, & \text { for } r \rightarrow+\infty,\end{cases} \\
	& R_{a s y}^{\mathrm{up}} \rightarrow \begin{cases}C_{\ell m \omega}^{\mathrm{up}} e^{i \omega r^*}+\left(\Delta^0\right)^2 C_{\ell m \omega}^{\mathrm{ref}} e^{-i \omega r^*}, & \text { for } r \rightarrow r_{+}, \\
		C_{\ell m \omega}^{\operatorname{trans}} r^3 e^{i \omega r^*}, & \text { for } r \rightarrow+\infty,\end{cases}
\end{align}
with the tortoise coordinate $r^*$ defined by $r^*=\int \frac{r^2}{\Delta^0} d r$, and the inhomogeneous solution for the radial equation being
\begin{align}
	\nonumber R_{\ell m \omega}(r) & =\frac{1}{2 i \omega C_{\ell m \omega}^{\mathrm{trans}} B_{\ell m \omega}^{\mathrm{inc}}}\left\{R_{\ell m \omega}^{\mathrm{up}}(r) \int_{r_{+}}^r d \tilde{r} \frac{f(\tilde{r})R_{\ell m \omega}^{\mathrm{in}}(\tilde{r})T_{\ell m \omega}(\tilde{r})}{\left(\Delta^0\right)^2}\right. \\
	& \left.+R_{\ell m \omega}^{\mathrm{in}}(r) \int_r^{\infty} d \tilde{r} \frac{f(\tilde{r})R_{\ell m \omega}^{\mathrm{up}}(\tilde{r})T_{\ell m \omega}(\tilde{r})}{\left(\Delta^0\right)^2}\right\},
\end{align}
where $R_{\ell m \omega}^{\mathrm{up}}(\tilde{r})$ and $R_{\ell m \omega}^{\mathrm{in}}(\tilde{r})$ are homogeneous solutions which satisfy the outgoing-wave boundary condition at the infinity and the ingoing-wave boundary condition at the horizon, respectively, and $r_+$ denotes the radius of the event horizon. Therefore, the solution of the Teukolsky-like equation at infinity can be expressed as
\begin{align}\label{Zeq}
	R_{\ell m \omega}(r \rightarrow \infty)=\frac{r^3 e^{i \omega r^*}}{2 i \omega B_{\ell m \omega}^{\mathrm{inc}}} \int_{r_{+}}^{\infty} d \tilde{r} \frac{f(\tilde{r})R_{\ell m \omega}^{\mathrm{in}}(\tilde{r})T_{\ell m \omega}(\tilde{r})}{\left(\Delta^0\right)^2} \equiv \tilde{Z}_{\ell m \omega} r^3 e^{i \omega r^*}.
\end{align}
The above discussion shows that the key step to get $R_{\ell m \omega}(r \rightarrow \infty)$ is to find $R_{\ell m \omega}^{\mathrm{in}}(\tilde{r})$ that will be studied in the following.

\subsection{Analytical solution of homogeneous Teukolsky-like equation}

The Teukolsky-like equation without source can be rewritten as
\begin{align}
	\left[\frac{\Delta^2}{f(r)} \frac{d}{d r}\left(\frac{f(r)}{\Delta} \frac{d}{d r}\right)-V(r)\right] R_{l m \omega}=0.
\end{align}
It is difficult to solve the above equation directly owing to its long-range potential. However, we can transform it into a short-range potential S--N-like equation that takes the form
\begin{align}\label{sneq}
	\left[\frac{d^2}{d{r^{*}}^2}-\mathcal{F}(r)\frac{d}{dr^{*}}-\mathcal{U}(r)\right]X_{lm\omega}=0 ,
\end{align}
where
\begin{align}
	\nonumber	\mathcal{F}(r)&=\frac{\Delta}{r^2+a^2}\frac{\gamma^{\prime}}{\gamma},\\
	\nonumber	\mathcal{U}(r)&=\frac{\Delta U(r)}{r^2+a^2}^2+G(r)^2+\frac{\Delta}{r^2+a^2}G(r)^{\prime}-\frac{F(r)G(r)}{r^2+a^2},\\
	\nonumber	G(r)&=\frac{r\Delta}{(r^2+a^2)^{2}}+\frac{\Delta f(r)^{\prime}}{2f(r)(r^2+a^2)}-\frac{\Delta^{\prime}}{r^2+a^2},\\
	U(r)&=\frac{\Delta^2}{\beta}\left(\left(2\alpha+\frac{\beta^{\prime}}{\Delta}-\frac{\beta}{\Delta}\frac{f(r)^{\prime}}{f(r)}\right)^{\prime}-\frac{\gamma^{\prime}}{\gamma}\left(\alpha+\frac{\beta^{\prime}}{\Delta}-\frac{\beta}{\Delta}\frac{f(r)^{\prime}}{f(r)}\right)\right)+V(r) .
\end{align}
The relation between $ R_{l m \omega}$ and $X_{lm\omega}$ is
\begin{align}\label{traeq}
	R_{l m \omega}=\frac{1}{\gamma}\left[\left(\alpha+\frac{\beta^{\prime}}{\Delta}-\frac{f(r)^{\prime}\beta }{f(r)\Delta}\right)x(r)-\frac{x(r)^{\prime}\beta }{\Delta}\right] ,
\end{align}
with $x(r)=\frac{\Delta}{\sqrt{f(r)(r^2+a^2)}}X_{lm\omega}$, and the other functions that appear in the above equations are defined as
\begin{align}
	\nonumber	\alpha&=\frac{4iK}{r}+V(r)-\frac{K^2}{\Delta}+\frac{6\Delta}{r^2}-iK^{\prime}+\frac{iK\Delta^{\prime}}{\Delta},\\
	\nonumber	\beta&=\Delta\left(-2iK+\Delta\left(-\frac{4}{r}-\frac{f(r)^{\prime}}{f(r)}\right)+\Delta^{\prime}\right),\\	 \gamma&=\alpha\left(\alpha+\frac{\beta^{\prime}}{\Delta}-\frac{\beta}{\Delta}\frac{f(r)^{\prime}}{f(r)}-\frac{\beta}{\Delta}\left(\alpha^{\prime}+\frac{\beta}{\Delta^2}V(r)\right)\right) .
\end{align}
Then, the asymptotic behavior of the ingoing-wave solution $X^{in}_{lm\omega}$ is given by
\begin{align}
	& X_{asy}^{\mathrm{in}} \rightarrow \begin{cases}A_{lm\omega}^{\text{out}}e^{i \omega r^*}+A_{lm\omega}^{\text{in}}e^{-i \omega r^*}, & \text { for } r \rightarrow +\infty, \\
		C_{lm\omega}^{\text{trans}} r^3 e^{i \omega r^*}, & \text { for } r \rightarrow r_{+}.\end{cases}
\end{align}
Meanwhile, $A_{lm\omega}^{\text{in}}$ is related to $B_{lm\omega}^{\text{inc}}$ in Eq. (\ref{RBB}) as
\begin{align}\label{Blmin}
	B_{lm\omega}^{\text{inc}}=-\frac{1}{4\omega^2}A_{lm\omega}^{\text{in}} .
\end{align}

Now, we employ the PM expansion method for the effective spacetime background, as outlined in Refs.\cite{H1996,Jing2022}. Our focus is solely on the GWs emitted toward infinity; therefore, we need to solve for the ingoing-wave homogeneous Teukolsky-like function $R^{in}_{lm\omega}$, or its counterpart in the homogeneous S--N-like equation, namely $X^{in}_{lm\omega}$.

We first introduce a dimensionless variable $z$, a dimensionless parameter $\eta$, and a new parameter $q$ associated $a$ with $\eta$, which can be defined as
\begin{align}
	z\equiv\omega r, \quad \eta\equiv2GM\omega, \quad q\equiv \frac{2a\omega}{\eta} ,
\end{align}
and assume that the solution has the following form
\begin{align}\label{solform}
	X^{in}_{lm\omega}=\sqrt{z^2+\frac{\eta ^2 q^2}{4}}\xi_{lm}(z)e^{-i\phi(z)} ,
\end{align}
where the exponential term is intended to eliminate the singularity at the horizon and can be written as
\begin{align}
	\nonumber \phi(z) & = \int \left( \frac{(r^{2} + a^{2}) \omega - m a}{\Delta} - \omega\right)d r \\
	& = \eta \left(b_{1} ln \left(z - c_{1} \eta\right) + b_{2} ln \left(z - c_{2} \eta\right)- b_{h} ln \left(z - c_{h} \eta\right)\right) ,
\end{align}
with
\begin{align}
	c_{1} = & \frac{1}{3} - \frac{1}{2} \Big[ (1 - i \sqrt{3}) (Q + \sqrt{P^{3} + Q^{2}})^{\frac{1}{3}} + (1 + i \sqrt{3}) (Q - \sqrt{P^{3} + Q^{2}})^{\frac{1}{3}} \Big] , \\
	c_{2} = & \frac{1}{3} - \frac{1}{2} \Big[ (1 + i \sqrt{3}) (Q + \sqrt{P^{3} + Q^{2}})^{\frac{1}{3}} + (1 - i \sqrt{3}) (Q - \sqrt{P^{3} + Q^{2}})^{\frac{1}{3}} \Big] , \\
	c_{h} = & \frac{1}{3} + (Q + \sqrt{P^{3} + Q^{2}})^{\frac{1}{3}} + (Q - \sqrt{P^{3} + Q^{2}})^{\frac{1}{3}} , \\
	b_{1} = & \frac{c_{1}^{3}}{(c_{1} - c_{2})(c_{1} - c_{h})} \left( -\frac{m q}{2 c_{1}^2}+\frac{\eta  q^2}{4 c_{1}^2}+\eta \right) , \\
	b_{2} = & \frac{c_{2}^{3}}{(c_{2} - c_{1})(c_{2} - c_{h})} \left( -\frac{m q}{2 c_{2}^2}+\frac{\eta  q^2}{4 c_{2}^2}+\eta \right) , \\
	b_{h} = & \frac{c_{h}^{3}}{(c_{1} - c_{h})(c_{h} - c_{2})} \left( -\frac{m q}{2 c_{h}^2}+\frac{\eta  q^2}{4 c_{h}^2}+\eta \right) ,
\end{align}
where $Q = \frac{1}{27} - \frac{a_{2}+q^{2}}{24} - \frac{a_{3}}{16}$ and $P = \frac{1}{3} \Big( \frac{a_{2}+q^{2}}{4} - \frac{1}{3} \Big)$.

By inserting Eq.~($\ref{solform}$) into Eq.~($\ref{sneq}$) and expanding it to the third order in powers of $\eta$, we obtain
\begin{align}
	L^{(0)}\left[\xi_{lm}\right]=\eta\, L^{(1)}\left[\xi_{lm}\right]+\eta^{2} L^{(2)}\left[\xi_{lm}\right]+\eta^{3} L^{(3)}\left[\xi_{lm}\right],
\end{align}
where the specific form of these differential operators is given by
\begin{align}
	L^{(0)}&=\frac{d^2}{dz^2}+\frac{2}{z}\frac{d}{dz}+\left(1-\frac{\ell(\ell+1)}{z^2}\right),\\
	 L^{(1)} & =\frac{1}{z} \frac{d^2}{d z^2}+\left(\frac{1+2 i z}{z^2}+\frac{2a_2}{3z^2}\right)\frac{d}{d z}-\frac{4+z^2-i z}{z^3}-\frac{a_2\left(\ell^2+\ell+2\right)}{3z^3}+\mathfrak{Q}^{(1)}, \\
	L^{(2)} & =-\frac{a_2+q^2}{4 z^2} \frac{d^2}{d z^2}+\left(\frac{\mathfrak{a}_{\ell}^{(2)}}{z^2} +\frac{\mathfrak{b}_{\ell}^{(2)}}{z^3}\right) \frac{d}{d z}+\left(\frac{\mathfrak{c}_{\ell}^{(2)}}{z^2}+\frac{\mathfrak{d}_{\ell}^{(2)}}{z^3} +\frac{\mathfrak{e}_{\ell}^{(2)}}{z^4}\right), \\
    L^{(3)} & =-\frac{a_3}{8 z^3} \frac{d^2}{d z^2}+\left(\frac{\mathfrak{a}_{\ell}^{(3)}}{z^2}+\frac{\mathfrak{b}_{\ell}^{(3)}}{z^3} +\frac{\mathfrak{c}_{\ell}^{(3)}}{z^4}\right) \frac{d}{d z}+\left(\frac{\mathfrak{d}_{\ell}^{(3)}}{z^2} +\frac{\mathfrak{e}_{\ell}^{(3)}}{z^3}+\frac{\mathfrak{f}_{\ell}^{(3)}}{z^4} +\frac{\mathfrak{g}_{\ell}^{(3)}}{z^5}\right),
\end{align}
with
\begin{align}
	\mathfrak{Q}^{(1)}=-\frac{i m q\left(4+\ell+\ell^2\right)}{z^2 \ell(\ell+1)}\frac{d}{dz}-\frac{4imq}{z^3\ell\left(\ell+1\right)}
\end{align}
and the coefficients that appear in $L^{(2)}$ and $L^{(3)}$ can be found in Appendix $\ref{ApenB}$. Since the general expressions are too long, we only show the results when $\ell$ is fixed to 2.

To maintain the consistency of the perturbation, we also expand $\xi_{lm}$ with respect to $\eta$ as
\begin{align}
	\xi_{\ell m}=\sum_{n=0}^{\infty}\eta^{n}\xi_{\ell m}^{(n)}(z) .
\end{align}
At this point, we can solve the equation order by order. The recursive equation is
\begin{align}\label{receq}
	L^{(0)}\left[\xi^{(n)}_{\ell}\right]=W^{(n)}_{\ell} ,
\end{align}
where
\begin{align}\label{receq1}
	\nonumber W^{(0)}_{\ell m}&=0,\\
	\nonumber W^{(1)}_{\ell m}&=\left(L^{(1)}\right)\left[\xi^{(0)}_{\ell m}\right],\\
	\nonumber W^{(2)}_{\ell m}&=\left(L^{(1)}\right)\left[\xi^{(1)}_{\ell m}\right]+\left(L^{(2)}\right)\left[\xi^{(0)}_{\ell m}\right] ,\\
	W^{(3)}_{\ell m}&=\left(L^{(1)}\right)\left[\xi^{(2)}_{\ell m}\right]+\left(L^{(2)}\right)\left[\xi^{(1)}_{\ell m}\right]+\left(L^{(3)}\right)\left[\xi^{(0)}_{\ell m}\right] .
\end{align}

The solution to Eq.($\ref{receq}$) for the case $n=0$ can be expressed as a linear combination of two classes of the usual Bessel functions, i.e.
\begin{align}
	\xi^{(0)}_{\ell m}=\alpha^{(0)}_{\ell m}j_{\ell}+\beta^{(0)}_{\ell m}j_{\ell} .
\end{align}
To ensure that $\xi^{(n)}_{\ell m}$ meets the boundary condition of regularity at $z=0$ for $n\leq2$, and without loss of generality, we assign $\alpha^{(0)}_{\ell m}=1$ and $\beta^{(0)}_{\ell m}=0$.
For further calculations when $n\geq1$, Eq.~($\ref{receq1}$) should be rewritten in the indefinite integral form by using the spherical Bessel functions
\begin{align}\label{sbeq}
	\xi^{(n)}_{\ell m}=n_{\ell}\int^{z}dzz^2j_{\ell}W^{(n)}_{\ell m}-j_{\ell}\int^{z}dzz^2n_{\ell}W^{(n)}_{\ell m}\quad(n=1,2) .
\end{align}
Thus, the solution of Eq.~($\ref{solform}$) corresponds to the case of $n = 1$
is given by
\begin{align}
	\nonumber	\xi_{\ell m}^{(1)}= & \alpha_{\ell}^{(1)} j_{\ell}+\frac{(\ell-1)(\ell+3)}{2(\ell+1)(2 \ell+1)} j_{\ell+1}-\frac{\ell^2-4}{2 \ell(2 \ell+1)} j_{\ell-1}+\sum_{k=1}^{\ell-1}\left(\frac{1}{k}+\frac{1}{k+1}\right) z^2\left(n_{\ell }j_k-j_{\ell} n_k\right) j_k\\
	\nonumber	&+z^2\left(n_{\ell }j_0-j_{\ell} n_0\right) j_0-2\mathfrak{D}^{nj}_{\ell}-\frac{a_2}{3}\left(\frac{\ell^2+3\ell+4}{2\left(\ell+1\right)\left(2\ell+1\right)}j_{\ell+1}-\frac{\ell^2-\ell+2}{2\ell\left(2\ell+1\right)}j_{\ell-1}\right)\\
	&+i j_{\ell} ln z+\frac{i m q}{2}\left(\frac{\ell^2+4}{\ell^2(2 \ell+1)}\right) j_{\ell-1}+\frac{i m q}{2}\left(\frac{(\ell+1)^2+4}{(\ell+1)^2(2 \ell+1)}\right) j_{\ell+1},
\end{align}
with
\begin{align}
	\nonumber \mathfrak{D}^{nJ}_{\ell}=n_{\ell}B_{jJ}-j_{\ell}B_{nJ}&=n_{\ell}\int^{z}_{0}zj_0\mathfrak{D}^{J}_0dz-j_{\ell}\int^{z}_{0}zn_0\mathfrak{D}^{J}_0dz,\\
	\nonumber \mathfrak{D}^{jJ}_{\ell}=j_{\ell}B_{jJ}+n_{\ell}B_{nJ}&=j_{\ell}\int^{z}_{0}zj_0\mathfrak{D}^{J}_0dz+n_{\ell}\int^{z}_{0}zn_0\mathfrak{D}^{J}_0dz,\\
	\mathfrak{D}^{j}_\ell&\equiv j_\ell,\quad\mathfrak{D}^{n}_\ell\equiv n_\ell ,
\end{align}
and $\alpha^{(1)}_{\ell}$ is the integration constant which represents the arbitrariness of the normalization of $X^{in}_{\ell m \omega}$, for convenience, we set $\alpha^{(1)}_{\ell}=0$.

Owing to the complexity of the general expressions, we choose to present results specifically for particular values of $\ell$ concerning the second- and third-order terms of $\xi_{\ell m}$ by using Eq.~($\ref{sbeq}$). Specifically, we provide results for $\ell=2,3$ at the second order and for $\ell=2$ at the third order. Inserting these expressions into Eq.~($\ref{solform}$) and expanding it in terms of $\eta$, we can obtain the solution $X^{in}_{2 m \omega}$ of the S--N-like equation $(\ref{sneq})$. Furthermore, using the transformation given by Eq.~($\ref{traeq}$), we obtain
the corresponding solutions $R^{(in)}_{2 m \omega}$ of the Teukolsky-like equation without a source (see Appendix $\ref{ApenC}$ for explicit expressions of the aforementioned results).

At last, we should consider the amplitude $A^{in}_{\ell m \omega}$. By introducing the first and second kinds of spherical Hankel functions defined as
\begin{align}
	\nonumber	&h_{\ell}^{(1)}=j_{\ell}+i n_{\ell} \rightarrow(-1)^{\ell+1} \frac{e^{i z}}{z},\\
	&h_{\ell}^{(2)}=j_{\ell}-i n_{\ell} \rightarrow(-1)^{\ell+1} \frac{e^{-i z}}{z} ,
\end{align}
the spherical Bessel function can be expressed in terms of the two Hankel functions as
\begin{align}
	\nonumber	j_\ell&=\frac{1}{2}\left(h^{(1)}_\ell+h^{(2)}_\ell\right),\\
	n_\ell&=\frac{1}{2i}\left(h^{(1)}_\ell-h^{(2)}_\ell\right) .
\end{align}
Then examining the asymptotic behavior of $\xi^{(n)}_{\ell m}$ at $z\rightarrow\infty$, we obtain
\begin{align}\label{eq:amplitude}
	\nonumber A_{\ell m}^{in} = & \frac{1}{2} i^{\ell+1} e^{-i \eta  (ln 2 \eta + \mathbf{elg} )} e^{i \big[\eta  p_{\ell m}^{(0)} -\pi  \eta ^2 p_{\ell m}^{(1)} + \eta ^3 \big(p_{\ell m}^{(2)}-\pi ^2 p_{\ell m}^{(3)}+ p_{\ell m}^{(4)} \mathbf{RiZ}(3) \big)\big]} \bigg\{ 1 - \frac{\pi}{2} \eta + \\
	& \eta ^2 \Big[ 2 ( \mathbf{elg} +ln 2) p_{\ell m}^{(1)}+q_{\ell m}^{(1)}+\frac{5 \pi ^2}{24} \Big] + \eta ^3 \Big[ \pi  q_{\ell m}^{(2)}+\pi ^3 q_{\ell m}^{(4)}+\pi  ( \mathbf{elg} +ln 2) q_{\ell m}^{(3)} \Big] \bigg\}.
\end{align}
where $\mathbf{elg}$ is EulerGamma constant $\mathbf{elg} = 0.57721 \cdots$, $\mathbf{RiZ}(n)$ is the Riemann zeta function and $\mathbf{RiZ}(3) = 1.202 \cdots$, 
and the coefficients of $A_{2m}^{in}$ are
\begin{align}
	p_{2m}^{(0)} = & \frac{5}{3} - \frac{2 a_{2}}{9} - i \frac{m q}{18} , \quad p_{2m}^{(1)} = \frac{8 a_{2}^2}{945}-\frac{a_{2}}{42}+\frac{5 a_{3}}{252}+\frac{m^2 q^2}{945}-\frac{q^2}{105}+\frac{107}{420} , \\
	\nonumber p_{2m}^{(2)} = & \frac{109 a_{2}^2}{1944}-\frac{85 a_{2} a_{3}}{3888}-\frac{a_{2} q^{2} \mathcal{H}_{(2m)}}{288} -\frac{569 a_{2} m^2 q^2}{54432}+\frac{17 a_{2} q^2}{1134}-\frac{11 a_{2}}{216}+\frac{11 a_{3}}{108}-\frac{11 m^2 q^2}{3888}+\frac{q^2}{288} + \\
	& \frac{29}{648} - \frac{73 a_{2}^3}{8748} - i \left(\frac{a_{2}^2 m q}{1944}+\frac{5 a_{2} m q}{972} + \frac{115 a_{3} m q}{15552} + \frac{197 m q}{864}\right) , \\
	p_{2m}^{(3)} = & \frac{107}{1260} + \frac{8 a_{2}^2}{2835}-\frac{a_{2}}{126}+\frac{5 a_{3}}{756}+\frac{m^2 q^2}{2835}-\frac{q^2}{315} , \quad p_{2m}^{(4)} = \frac{1}{3} , \\
	q_{2m}^{(1)} = & \frac{25}{18} + \frac{q^2}{72} - \frac{37 a_{2}}{108}-\frac{5 a_{3}}{108}-\frac{m^2 q^2}{216} + i \left(\frac{2 a_{2} m q}{81}+\frac{5 m q}{108}\right) , \\
	\nonumber q_{2m}^{(2)} = & -\frac{25}{36} -\frac{a_{2}^2}{210}-\frac{5 a_{2} a_{3}}{378}-\frac{a_{2} q^2 \mathcal{H}_{(2m)} }{2520}-\frac{19 a_{2} m^2 q^2}{52920}+\frac{5 a_{2} q^2}{1764}+\frac{37 a_{2}}{216}+\frac{43 a_{3}}{3024}+\frac{m^2 q^2}{432}-\frac{q^2}{144} \\
	& -\frac{11 a_{2}^3}{2835} + i \left(\frac{a_{2}^2 m q}{270} -\frac{a_{2} m q}{2268}+\frac{17 a_{3} m q}{3024}+\frac{m q}{54}\right)  , \\
	\nonumber q_{2m}^{(3)} = & - \frac{107}{420} - \frac{a_{2} m q}{42 \pi }+\frac{a_{2}}{42}-\frac{17 a_{3} m q}{1512 \pi }-\frac{5 a_{3}}{252}-\frac{m^2 q^2}{945}-\frac{m q}{12 \pi }+\frac{q^2}{105} -\frac{a_{2}^2 m q}{135 \pi }-\frac{8 a_{2}^2}{945} + \\
	& i \left( \frac{5 a_{2} q^2}{882 \pi } - \frac{22 a_{2}^3}{2835 \pi } - \frac{a_{2}^2}{105 \pi } - \frac{5 a_{2} a_{3}}{189 \pi } - \frac{a_{2} q^{2} \mathcal{H}_{(2m)}}{1260 \pi }-\frac{19 a_{2} m^2 q^2}{26460 \pi } - \frac{a_{3}}{56 \pi } \right) , \\
	q_{2m}^{(4)} = & -\frac{1}{16} .
\end{align}

\section{Analytical solution of \texorpdfstring{ $\psi_4^{B}$}{} with source in effective spacetime}\label{sec:4}
\subsection{Tetrad components of energy--momentum tensor}
The energy--momentum tensor for the EOB theory, which describes a particle orbiting a massive black hole described by the effective metric, can be expressed as
\begin{align}
	T^{\mu \nu}=\frac{m_0}{\Sigma \sin \theta d t / d \tau} \frac{d x^\mu}{d \tau} \frac{d x^\nu}{d \tau} \delta(r-r(t)) \delta(\theta-\theta(t)) \delta(\varphi-\varphi(t)) ,
\end{align}
where $m_0$ is the mass of the test particle, $x^\mu=(t, r(t), \theta(t), \varphi(t))$ is a geodesic trajectory, and $\tau$ is the proper time along the geodesic. The geodesic equations in the effective metric are
\begin{align}
	\nonumber & \Sigma \frac{d \theta}{d \tau}= \pm\left[\mathscr{C}-\cos ^2 \theta\left\{a^2\left(1-E^2\right)+\frac{L^2}{\sin ^2 \theta}\right\}\right]^{1 / 2}, \\
	\nonumber & \Sigma \frac{d \varphi}{d \tau}=-\left(a E-\frac{L}{\sin ^2 \theta}\right)+\frac{a}{\Delta}\left(E\left(r^2+a^2\right)-a L\right), \\
	\nonumber & \Sigma \frac{d t}{d \tau}=-\left(a E-\frac{L}{\sin ^2 \theta}\right) a \sin ^2 \theta+\frac{r^2+a^2}{\Delta}\left(E\left(r^2+a^2\right)-a L\right), \\
	& \Sigma \frac{d r}{d \tau}= \pm\left[E\left(r^2+a^2\right)-a L\right]^2-\Delta\left[(E a-L)^2+r^2+\mathscr{C}\right]^{1 / 2},
\end{align}
where $E$, $L$ and $\mathscr{C}$ are the energy, the $z$-component of the angular momentum, and the Carter constant of a test particle, respectively.

Thus, the tetrad components of the energy--momentum tensor can be expressed as
\begin{align}
	\nonumber T_{n n} & =m_0 \frac{C_{n n}}{\sin \theta} \delta(r-r(t)) \delta(\theta-\theta(t)) \delta(\varphi-\varphi(t)), \\
	\nonumber T_{\overline{m} n} & =m_0 \frac{C_{\overline{m} n}}{\sin \theta} \delta(r-r(t)) \delta(\theta-\theta(t)) \delta(\varphi-\varphi(t)), \\
	T_{\overline{m m}} & =m_0 \frac{C_{\overline{m m}}}{\sin \theta} \delta(r-r(t)) \delta(\theta-\theta(t)) \delta(\varphi-\varphi(t)),
\end{align}
with
\begin{align}
	\nonumber C_{n n} & =\frac{1}{4 \Sigma^3 \dot{t}}\left[E\left(r^2+a^2\right)-a L+\Sigma \frac{d r}{d \tau}\right]^2, \\
	\nonumber C_{\overline{m} n} & =-\frac{1}{2 \sqrt{2} \bar{\rho}^* \Sigma^2 \dot{t}}\left[E\left(r^2+a^2\right)-a L+\Sigma \frac{d r}{d \tau}\right]\left[i \sin \theta\left(a E-\frac{L}{\sin ^2 \theta}\right)+\Sigma \frac{d \theta}{d \tau}\right] ,\\
	C_{\overline{m m}} & =\frac{1}{2 \bar{\rho}^{* 2} \Sigma \dot{t}}\left[i \sin \theta\left(a E-\frac{L}{\sin ^2 \theta}\right)+\Sigma \frac{d \theta}{d \tau}\right]^2,
\end{align}
where $\dot{t}=d t / d \tau$.
Now we expand equation (3.30) of~\cite{Jing2023} for $T_{\ell m \omega}(r)$ to second order in the rotation parameter $a$, i.e.
\begin{align}\label{Teq}
	T_{\ell m \omega}(r)=T_{\ell m \omega}^{(0)}(r)+a T_{\ell m \omega}^{(1)}(r)+a^2 T_{\ell m \omega}^{(2)}(r),
\end{align}
with
\begin{align}
	\nonumber & T_{\ell m \omega}^{(0)}(r)=\frac{1}{2 \pi} \int_{-\infty}^{+\infty} d t \int d \Omega \mathcal{T}_4^{(0)} e^{i(\omega t-m \varphi)} \frac{{}_{-2} Y_{\ell m}^*(\theta)}{\sqrt{2 \pi}} ,\\
	\nonumber & T_{\ell m \omega}^{(1)}(r)=\frac{1}{2 \pi} \int_{-\infty}^{+\infty} d t \int d \Omega \mathcal{T}_4^{(1)} e^{i(\omega t-m \varphi)} \frac{{}_{-2} Y_{\ell m}^*(\theta)}{\sqrt{2 \pi}} ,\\
	& T_{\ell m \omega}^{(2)}(r)=\frac{1}{2 \pi} \int_{-\infty}^{+\infty} d t \int d \Omega \mathcal{T}_4^{(2)} e^{i(\omega t-m \varphi)} \frac{{}_{-2} Y_{\ell m}^*(\theta)}{\sqrt{2 \pi}} ,
\end{align}
For a source bounded in a finite range of $r$, it is convenient to rewrite the source as
\begin{align}
	\nonumber T_{\ell m \omega}^{(0)}(r)= & -m_0 G \int_{-\infty}^{\infty} d t e^{i \omega t-i m \varphi(t)}\left(\Delta^0\right)^2\left\{\left(A_{n n 0}^{(0)}+A_{\overline{m} n 0}^{(0)}+A_{m m 0}^{(0)}\right) \delta(r-r(t))\right. \\
	& \left.+\left[\left(A_{\overline{m} n1}^{(0)}+A_{m m 1}^{(0)}\right) \delta(r-r(t))\right]^{\prime}+\left[A_{\overline{m m} 2}^{(0)} \delta(r-r(t))\right]^{\prime \prime}\right\} ,\\
	\nonumber T_{\ell m \omega}^{(1)}(r)= & -m_0 G \int_{-\infty}^{\infty} d t e^{i \omega t-i m \varphi(t)}\left(\Delta^0\right)^2\left\{\left(A_{n n 0}^{(1)}+A_{\overline{m} n 0}^{(1)}+A_{\overline{m m} 0}^{(1)}\right) \delta(r-r(t))\right. \\
	& \left.+\left[\left(A_{\overline{m} n 1}^{(1)}+A_{\overline{m m} 1}^{(1)}\right) \delta(r-r(t))\right]^{\prime}+\left[A_{\overline{m m} 2}^{(1)} \delta(r-r(t))\right]^{\prime \prime}\right\},\\
	\nonumber T_{\ell m \omega}^{(2)}(r)= & -m_0 G \int_{-\infty}^{\infty} d t e^{i \omega t-i m \varphi(t)}\left(\Delta^0\right)^2\left\{\left(A_{n n 0}^{(2)}+A_{\overline{m} n 0}^{(2)}+A_{\overline{m m} 0}^{(2)}\right) \delta(r-r(t))\right. \\
	& \left.+\left[\left(A_{\overline{m} n 1}^{(2)}+A_{\overline{m m} 1}^{(2)}\right) \delta(r-r(t))\right]^{\prime}+\left[A_{\overline{m m} 2}^{(2)} \delta(r-r(t))\right]^{\prime \prime}\right\},
\end{align}
and the explicit expressions of $A^{i}_{nn0}$, $A^{i}_{\overline{m}n0}$, ${A^{i}}_{\overline{mm}0}$, $A^{i}_{\overline{m}n1}$, ${A^{i}}_{\overline{mm}1}$ and ${A^{i}}_{\overline{mm}2}$ are given in Appendix $\ref{ApenD}$.

\subsection{Analytical solution of \texorpdfstring{ $\psi_4^{B}$}{} with source in effective spacetime}

Inserting Eqs. (\ref{Blmin}), the solution $R^{in}_{\ell m \omega}$ and (\ref{Teq}) into Eq.~(\ref{Zeq}), we obtain $\tilde{Z}_{\ell m \omega}$ as
\begin{align}
	\tilde{Z}_{\ell m \omega}=\sum_n \delta\left(\omega-\omega_n\right) Z_{\ell m \omega},
\end{align}
with
\begin{align}\label{zeq}
	 Z_{\ell m \omega}=\frac{\pi \nu GM}{i \omega B_{\ell m \omega}^{\mathrm{inc}}}\left[A_0 f(r)R_{\ell m \omega}^{\mathrm{in}}(r)-A_1\left(f(r)R_{\ell m \omega}^{\mathrm{in}}(r)\right)^{\prime}\right.\left.+A_2\left(f(r)R_{\ell m \omega}^{\mathrm{in}}(r)\right)^{\prime \prime}\right]_{r_0, \theta_0} ,
\end{align}
where $\omega_n=m\, \Omega$, $A_0=A_{n n 0}+A_{\overline{m} n 0}+A_{\overline{m m} 0}$, $A_1=A_{\overline{m} n 1}+A_{\overline{m m} 1}$, $A_2=A_{\overline{m m} 2}$ (see Appendix $\ref{ApenD}$ for details), and $\left(r_0, \theta_0\right)$ are the values of $(r, \theta)$ on the geodesic trajectory. Formula ($\ref{zeq}$) can also be rewritten in another intuitive form that is given by
\begin{align}
	Z_{\ell m \omega}=Z_{\ell m \omega}^{(NS)}+Z_{\ell m \omega}^{(q)}+ Z_{\ell m \omega}^{(q^2)}.
\end{align}
Where $Z_{\ell m \omega}^{(NS)}$ represents the part that does not relate to spin. That is, $Z_{\ell m \omega}$ can also be written as the expansion of the rotation parameter $a$, when $a\rightarrow 0$, the result will degenerate into the nonspinning case~\cite{L2023}.

Then, in particular, $\psi_4^{B}$ at $r\rightarrow \infty$ is obtained from Eq.($\ref{psieq}$) as
\begin{align}\label{psi4}
	\psi_4^{B} =\frac{1}{r} \sum_{\ell m n}Z_{\ell m \omega_n} \frac{{}_{-2}{Y}_{\ell m}}{\sqrt{2 \pi}} e^{i \omega_n\left(r^*-t\right)+i m \varphi}.
\end{align}

\section{Energy flux and waveform for plus and cross modes of gravitational wave}\label{sec:5}
Based on Eq. (\ref{psi4}), we find that the energy flux of the gravitational radiation can be described by
\begin{align}\label{enegyflux}
	\frac{d E}{d t} & =\int \frac{1}{16 \pi G}\left(\dot{h}_{+}^2+\dot{h}_{\times}^2\right) r^2 d \Omega =\sum_{\ell=2}^{\infty} \sum_{m=1}^{\ell} \frac{\left|Z_{\ell m \omega_n}\right|^2}{2 \pi G \omega_n^2}.
\end{align}
The reduced RRF is given by~\cite{Buo2006}
\begin{align}
	\hat{\boldsymbol{\mathcal{F}}}=\frac{1}{\nu M \Omega|\boldsymbol{r} \times \boldsymbol{P}|} \frac{d E}{d t} \boldsymbol{P},
\end{align}
where $\nu$ is the symmetric mass ratio, $\boldsymbol{P}$ represents the momentum vector of the effective particle, and $\Omega=|\boldsymbol{r} \times \dot{\boldsymbol{r}}| / r^2$ denotes the dimensionless orbital frequency.
For the quasi-circular case without precession, noting that $\mid \boldsymbol{r} \times$ $\boldsymbol{P} \mid \approx p_{\varphi}$, so the reduced RRF can be explicitly expressed as
\begin{align}\label{FFF}
	\hat{\mathcal{F}} =\frac{1}{\nu M \Omega} \sum_{\ell=2}^{\infty} \sum_{m=1}^{\ell} \frac{\left|Z_{\ell m \omega_n}\right|^2}{2 \pi G \omega_n^2}\frac{\boldsymbol{P}}{p_{\varphi}}.
\end{align}

On the other hand, it is well known that the plus and cross modes of the GW can be expressed in terms of spin-weighted $s = -2$ spherical harmonics~\cite{Kid2008}
\begin{align}\label{heq}
	h_{+}-i h_{\times}=\sum_{l=2}^{\infty} \sum_{m=-l}^l h^{l m} \frac{{}_{-2} Y_{l m}}{\sqrt{2 \pi}}e^{i \omega_n\left(r^*-t\right)+i m \varphi}.
\end{align}
Therefore, by comparing Eq. ($\ref{heq}$) with the solution (\ref{psi4}) of $\psi_4^{B}$, we can easily deduce the waveform,
which is given by
\begin{align}\label{HHH}
	h_{\ell m} & =-\frac{2}{r \omega_{n}^2}Z_{\ell m \omega_n}=h_{\ell m}^{(N,\epsilon)}\hat{h}_{\ell m}.
\end{align}
where $h_{\ell m}^{(N,\epsilon)}$ represents the Newtonian contribution, and $\epsilon$ denotes the parity of the multipolar waveform.

The next two figures seek to demonstrate a comparative analysis of our calculation results with those from other works, specifically for $\hat{h}_{\ell m}$ of GWs at a given $``r"$ that does not evolve over time.

\begin{figure*}[h]
	\centering
	\includegraphics[width=0.95\textwidth]{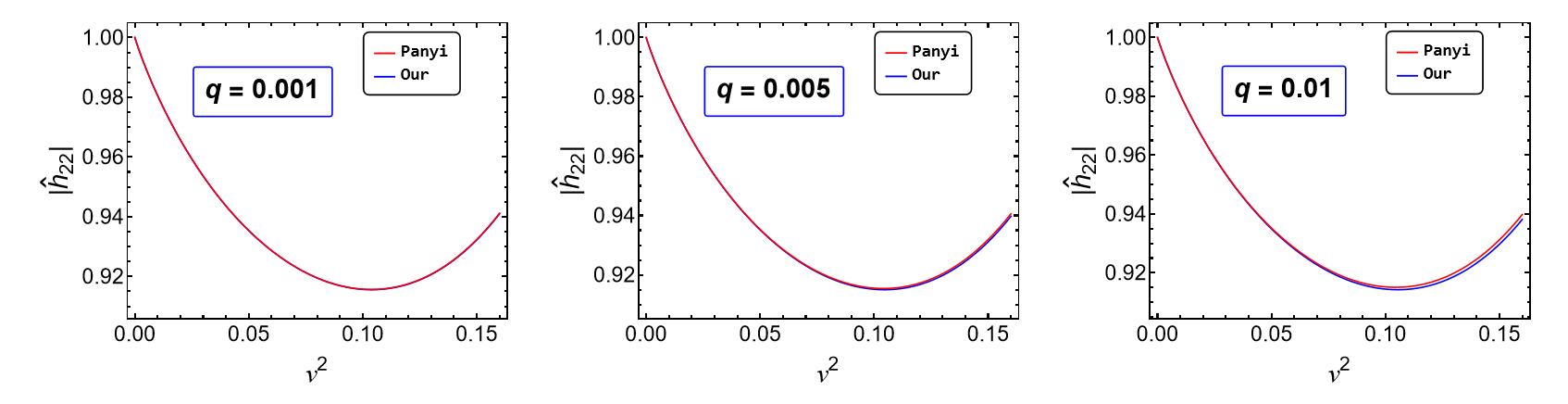}
	\caption{These images depict the comparative analysis of the curves of $\hat{h}_{22}$ conducted independently by this article and ref~\cite{Pan20111}. The values in the three pages from left to right are $q=0.001, 0.005, 0.01$, respectively.}
	\label{fig:1}
\end{figure*}

\begin{figure*}[h]
	\centering
	\includegraphics[width=0.95\textwidth]{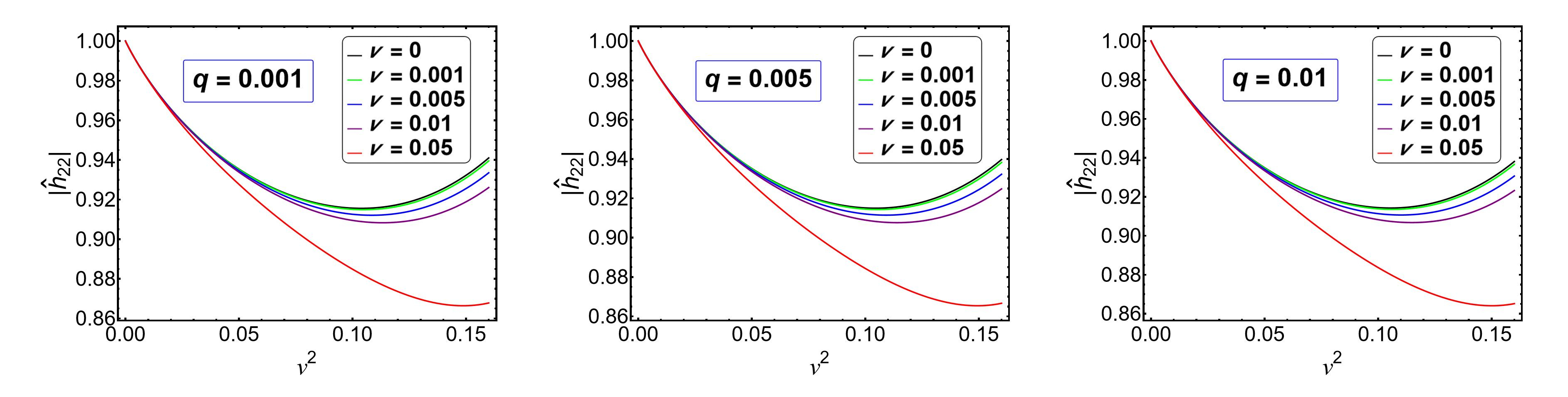}
	\caption{These images show the curves of $\hat{h}_{22}$ across five distinct symmetric mass ratios, with parameter $q$ ranging from 0.001 to 0.01.}
	\label{fig:2}
\end{figure*}

In Figure~\ref{fig:1}, we present the curves of the dominant $\hat{h}_{22}(q, \nu, v^2)$ mode as the parameter $q$ takes three different values, demonstrating that our results are essentially consistent with Panyi's~
\cite{Pan20111}. While Figure~\ref{fig:2} shows the curves of $\hat{h}_{22}(q, \nu, v^2)$ for different symmetric mass ratios $\nu$. Both achieve accuracy up to $v^9$ (where we define $v= \sqrt[3]{GM\Omega}$), with our results aligning with the Schwarzschild case at $v^9$ when setting the correction parameters and rotation parameter to zero.

Since $Z_{\ell m \omega}$ is obtained in the effective metric ($\ref{metric}$), Eqs. (\ref{enegyflux}),  (\ref{FFF}) and (\ref{HHH}) indicate that the energy flux, reduced RRF, and the waveform are constructed in terms of effective spacetime.

\section{Conclusions}\label{sec:6}
It is known that nonspinning binary systems are largely theoretical constructs, as real-world conditions involve rotation to some extent. From a practical perspective, it is essential to develop gravitational waveform templates that account for spin binaries. Therefore, as a theory for building a template of gravitational waveforms, it is particularly important to consider the case of spinning binaries.

To construct the gravitational waveform template for the radiation generated by merging compact object binary systems, we must focus on the late dynamical evolution of the system governed by the Hamilton equation. This involves analyzing the Hamiltonian and the RRF of the binary system. Based on previous work~\cite{Jing2022,Jing20231}, we have built up an effective rotating metric for a real spinning two-body system and constructed an improved SCEOB Hamiltonian. This Hamiltonian models a spinning test particle's orbit around a massive rotating black hole, as described by the aforementioned metric. The null tetrad component of the perturbed Weyl tensor $\psi^B_{4}$ relates to the plus and cross modes of the GWs as $\psi_{4}^{B}=\frac{1}{2}(\ddot{h}_{+}-i\ddot{h}_{\times})$ at infinity. This relationship is crucial for calculating the GW energy flux and, by extension, the RRF. The key step is to find the solution of $\psi^B_{4}$.

Although the decoupled equation \cite{Jing2023} initially appears inseparable, insights from analyzing the LIGO--Virgo strain data proposed by Renolds et al. reveal that most spin systems involve slowly spinning black holes, namely $a<0.1$. This enables us to separate the equation into radial and angular parts within a slowly rotating background, closely approximating spherical symmetry. Thus, we have further obtained the variable separation form of the decoupled equation up to the second order with respect to the rotation parameter $a$, and this method can, in principle, be extended to any order. By rewriting the trigonometric function in terms of the spin-weighted spherical harmonics and leveraging their characteristics, we have extracted an analytical solution for the radial equation through certain transformations. Utilizing the null tetrad component of the perturbed Weyl tensor $\psi_{4}^{B}$, we have derived analytical expressions for the reduced RRF and the waveform for the plus and cross modes of the GWs. At this point, the late-stage dynamic evolution of the binary system becomes clear and distinct.

It is worth noting that our results can be decomposed into several components. Aside from the basic Schwarzschild background, these include elements related to the rotation parameter $a$, the correction parameter $a_2$ and $a_3$, and their coupling terms. Specifically, when $a\rightarrow 0$, the result will be reduced to the case of~\cite{L2023}, while $a_2, a_3\rightarrow 0$, the result will be consistent with the Kerr background. If all these parameters approach 0, the model simplifies to the simplest Schwarzschild case. Meanwhile, the improved reduced EOB Hamiltonian presented in~\cite{Jing2023}, along with the reduced RRF and the waveform for the plus and cross modes of the GW, are all obtained within the same physical model. These components collectively form a SCEOB theory for spinning black-hole binaries based on the PM approximation.

Moving forward, we aim to extend these results to a higher order to improve the accuracy and facilitate comparisons with data produced by numerical relativistic methods. Concurrently, we plan to plot the corresponding waveforms.

\section*{Acknowledgement}
This work was supported by the Grant of NSFC No. 12035005,
and National Key Research and Development Program of China No. 2020YFC2201400.

\vspace{1cm}
\appendix
%\begin{appendices}

\section{Expressions for \texorpdfstring{ $F^{i}_{j}$}{} }\label{ApenA}

The expressions of the coefficients $F^{i}_{j}$ is given by
\begin{align}
	F^{0}_{1}=&-\frac{32 \Delta^0-23 r \Delta^{0^{\prime}}+7 r^2 \Delta^{0^{\prime \prime}}-r^3 \Delta^{0(3)}}{r\left[8 \Delta^0-r\left(5 \Delta^{0^{\prime}}-r \Delta^{0^{\prime \prime}}\right)\right]}, \\
	\nonumber F^{1}_{1}=&-\frac{3}{r^2\left[8 \Delta^0-r\left(5 \Delta^{0^{\prime}}-r \Delta^{0^{\prime \prime}}\right)\right]^2}\left\{32\left(\Delta^0\right)^2+4 r \Delta^0\left[r\left(4+r \Delta^{0(3)}\right)-10 \Delta^{0^{\prime}}\right]\right. \\
	&\left.+r^2\left[17\left(\Delta^{0^{\prime}}\right)^2+r^2\left(\Delta^{0^{\prime \prime}}\left(8+\Delta^{0^{\prime \prime}}\right)-2 r \Delta^{0(3)}\right)-r \Delta^{0^{\prime}}\left(16+6 \Delta^{0^{\prime \prime}}+r \Delta^{0(3)}\right)\right]\right\}, \\
	\nonumber F^{2}_{1}=& \frac{1}{r^3\left[8 \Delta^{0}+r\left(-5 \Delta^{0^{\prime}}+r \Delta^{0^{\prime \prime}}\right)\right]^3}\left\{2432 (\Delta^{0})^3+16 r (\Delta^{0})^2\left[-282 \Delta^{0^{\prime}}+r\left(54 \Delta^{0^{\prime \prime}}+r \Delta^{0^{\prime \prime\prime}}\right)\right]\right. \\
	\nonumber	& \left. +4 r^2 \Delta\left[702\left(\Delta^{0^{\prime}}\right)^2-r \Delta^{0^{\prime}}\left(60+246 \Delta^{0^{\prime \prime}}+17 r \Delta^{0^{\prime \prime\prime}}\right)+r^2\left(18\left(\Delta^{0^{\prime \prime}}\right)^2+8\left(9+2 r \Delta^{0^{\prime \prime\prime}}\right)\right. \right. \right. \\
	\nonumber	& \left. \left. \left. +\Delta^{0^{\prime \prime}}\left(-12+5 r \Delta^{0^{\prime \prime\prime}}\right)\right)\right] -r^3\left[647\left(\Delta^{0^{\prime}}\right)^3-r\left(\Delta^{0^{\prime}}\right)^2\left(312+381 \Delta^{0^{\prime \prime}}+16 r \Delta^{0^{\prime \prime\prime}}\right)\right.\right.\\
	\nonumber	& \left. \left. +r^3\left(-48\left(\Delta^{0^{\prime \prime}}\right)^2-7\left(\Delta^{0^{\prime \prime}}\right)^3+36 r \Delta^{0^{\prime \prime\prime}}+2 \Delta^{0^{\prime \prime}}\left(-72+5 r \Delta^{0^{\prime \prime\prime}}\right)\right)\right. \right. \\
	& \left.\left.+r^2 \Delta^{0^{\prime}}\left(288+81\left(\Delta^{0^{\prime \prime}}\right)^2-14 r \Delta^{0^{\prime \prime\prime}}+\Delta^{0^{\prime \prime}}\left(216+5 r \Delta^{0^{\prime \prime\prime}}\right)\right)\right]\right\},\\
	F^{0}_{2}=&\frac{8 \Delta^0-r\left(5 \Delta^{0^{\prime}}-r \Delta^{0^{\prime \prime}}\right)}{4 \Delta^0-r\left(2 r+\Delta^{0^{\prime}}\right)}, \\
	F^{1}_{2}=&-\frac{3}{r} \frac{4 \Delta^0-r\left[4 \Delta^{0^{\prime}}-r\left(2+\Delta^{0^{\prime \prime}}\right)\right]}{4 \Delta^0-r\left(2 r+\Delta^{0^{\prime}}\right)}, \\
	\nonumber F^{2}_{2}	=& \frac{1}{r^2\left[-4 (\Delta^{0})+r\left(2 r+\Delta^{0^{\prime}}\right)\right]^2}\left\{4\left(-24 (\Delta^{0})^2+r^2\left[-9\left(\Delta^{0^{\prime}}\right)^2+r^2\left(4+\Delta^{0^{\prime \prime}}\right)\right.\right.\right.\\
	&\left.\left.\left.+r \Delta^{0^{\prime}}\left(-1+2 \Delta^{0^{\prime \prime}}\right)\right]+r (\Delta^{0})\left[33 \Delta^{0^{\prime}}-r\left(8+5 \Delta^{0^{\prime \prime}}\right)\right]\right)\right\}, \\
	F^{2}_{3}=	& \frac{4 (\Delta^{0})+r\left(-14 r+5 \Delta^{0^{\prime}}\right)}{r^2\left[-4 (\Delta^{0})+r\left(2 r+\Delta^{0^{\prime}}\right)\right]}, \\
	F^{0}_{4}=&\frac{\left[8 \Delta^0-r\left(5 \Delta^{0^{\prime}}-r \Delta^{0^{\prime \prime}}\right)\right]}{2 r^4}, \\
	F^{1}_{4}=&\frac{3\left[4 \Delta^0-r\left(2 r+\Delta^{0^{\prime}}\right)\right]}{2 r^5} ,\\
	F^{2}_{4}	=& -\frac{28 (\Delta^{0})+r\left[-10 \Delta^{0^{\prime}}+r\left(-10+\Delta^{0^{\prime \prime}}\right)\right]}{2 r^6},\\
	F^{0}_{5}	=& \frac{4\Delta^0-r\left(2r+\Delta^{0^{\prime}}\right)}{2r^4},\\
	F^{1}_{5}	=& -\frac{3\left[-4\Delta^0+r\left(2r+\Delta^{0^{\prime}}\right)\right]}{2r^5},\\
	F^{2}_{5}	=& \frac{-10\Delta^0+r\left(8r+\Delta^{0^{\prime}}\right)}{r^6}.
\end{align}

\section{Expressions of operators \texorpdfstring{ $L^{(2)}$}{} and \texorpdfstring{ $L^{(3)}$}{} }\label{ApenB}

The expressions of the coefficients for the operator $L^{(2)}$ at $\ell=2$ are given by
\begin{align}
	\mathfrak{a}_{2}^{(2)}=&-\frac{ia_2}{6}-\frac{4ia_2^2}{27}-\frac{5ia_3}{18}+2i\left(c_1^2+\left(c_1+c_2\right)\left(c_2-1\right)\right)+\left(\frac{1}{6}-\frac{2a_2}{27}\right)mq+\left(\frac{7i}{12}-\frac{im^2}{54}\right)q^2,\\
	\mathfrak{b}_{2}^{(2)}=&-\frac{5a_2}{6}+\frac{8a_2^2}{9}+\frac{25a_3}{12}+\left(\frac{i}{6}-\frac{ia_2}{3}\right)mq+\left(-\frac{1}{4}+\frac{m^2}{18}\right)q^2,\\
	\mathfrak{c}_{2}^{(2)}=& \frac{5a_2}{12}+\frac{4a_2^2}{27}+\frac{5a_3}{18}+\left(\frac{i}{6}-\frac{2ia_2}{27}\right)mq+\left(-\frac{5}{42}-\frac{m^2}{189}\right)q^2,\\
	\mathfrak{d}_{2}^{(2)}=& -ia_2+\frac{8ia_2^2}{27}+\frac{5ia_3}{9}+\frac{4a_2 mq}{27}+\left(-\frac{i}{6}+\frac{im^2}{27}\right)q^2,\\
	\mathfrak{e}_{2}^{(2)}=& \frac{19a_2}{6}-\frac{19a_2^2}{9}-\frac{145a_3}{24}+\left(-\frac{i}{3}+\frac{35ia_2}{18}\right)mq+\left(1+\frac{5m^2}{9}\right)q^2.
\end{align}

While the coefficients of the operator $L^{(3)}$ at $\ell=2$ can be written as
\begin{align}
	\nonumber
	\mathfrak{a}_{2}^{(3)}=&\frac{a_2}{24}+\frac{a_2^2}{54}-\frac{4a_2^3}{243}+\frac{5a_3}{72}-\frac{5a_2 a_3}{162}+\left(\frac{i}{24}-\frac{ia_2 }{9}-\frac{8i a_2^2 }{243}-\frac{25i a_3 }{324}\right)mq,\\
	&+\left(-\frac{1}{48}+\frac{17m^2}{216}+\frac{a_2}{108}-\frac{10a_2 m^2}{243}\right)q^2\\
	\nonumber \mathfrak{b}_{2}^{(3)}=&\frac{5ia_2}{8}+\frac{ia_2^2}{18}-\frac{26ia_2^3}{81}-\frac{5ia_3}{36}-\frac{55ia_2 a_3}{54}+\left(-\frac{1}{24}-\frac{5a_2}{18}-\frac{17a_2^2 }{81}-\frac{55a_3 }{216}\right)mq\\
	&+\left(\frac{29i}{48}-\frac{65im^2}{216}-\frac{29ia_2}{252}+\frac{i a_2 \mathcal{H}_{(2m)}}{12}+\frac{407ia_2 m^2}{2268}\right)q^2-2i\left(c_1-1\right)\left(c_2-1\right)\left(c_1+c_2\right),\\
	\nonumber \mathfrak{c}_{2}^{(3)}=&\frac{5a_2}{12}-\frac{8a_2^2}{9}+\frac{10a_2^3}{9}-\frac{31a_3}{12}+\frac{145a_2 a_3}{36}+\left(-ic_1^2-i\left(c_1+c_2\right)\left(c_2-1\right)	\right.\\
	& \left.+\frac{i}{2}+\frac{ia_2}{3}-\frac{2ia_2^2}{3}-\frac{5ia_3}{6}\right)mq+\left(\frac{5}{8}-\frac{5m^2}{9}-\frac{47a_2}{84}-\frac{a_2 \mathcal{H}_{(2m)}}{12}+\frac{25a_2 m^2}{252}\right)q^2,\\
	\nonumber \mathfrak{d}_{2}^{(3)}=&\frac{ia_2}{24}+\frac{ia_2^2}{54}-\frac{4ia_2^3}{243}+\frac{5ia_3}{72}-\frac{5ia_2 a_3}{162}+\left(-\frac{1}{24}+\frac{a_2}{9}+\frac{8a_2^2}{243}+\frac{25a_3}{324}\right)mq\\
	&\left(-\frac{i}{48}+\frac{17im^2}{216}+\frac{ia_2}{108}-\frac{10ia_2 m^2}{243}\right)q^2,\\
	\nonumber \mathfrak{e}_{2}^{(3)}=&\frac{5a_2}{12}-\frac{a_2^2}{27}+\frac{74a_2^3}{243}+\frac{a_3}{3}+\frac{80a_2 a_3}{81}+2c_1^2+2\left(c_1+c_2\right)\left(c_2-1\right)+\left(-\frac{7ia_2}{18}-\frac{59ia_2^2}{243}	\right.\\
	& \left.-\frac{215ia^3}{648}\right)mq+\left(\frac{3}{8}+\frac{41m^2}{108}+\frac{10a_2}{189}-\frac{a_2 \mathcal{H}_{(2m)}}{12}-\frac{1447a_2 m^2}{6804}\right)q^2,\\
	\nonumber \mathfrak{f}_{2}^{(3)}=&\frac{ia_2}{6}-ia_2^2+\frac{16ia_2^3}{27}-\frac{55ia_3}{18}+\frac{35ia_2 a_3}{18}+i\left(c_1-1\right)\left(c_2-1\right)\left(c_1+c_2\right)-\frac{2ia_2}{3}c_1^2\\
	\nonumber&-\frac{2ia_2}{3}\left(c_1+c_2\right)\left(c_2-1\right)+\left(-\frac{5}{12}-\frac{55a_2}{36}+\frac{20a_2^2}{81}+\frac{25a_3}{108}-\frac{5}{3}c_1^2	\right.\\
	& \left.-\frac{5}{3}\left(c_1+c_2\right)\left(c_2-1\right)\right)mq+\left(\frac{13i}{24}-\frac{103im^2}{108}-\frac{53ia_2}{126}+\frac{52ia_2 m^2}{189}\right)q^2,\\
	\nonumber \mathfrak{g}_{2}^{(3)}=&\frac{7a_2}{6}+\frac{29a_2^2}{18}-\frac{52a_2^3}{27}+6a_3-\frac{70a_2 a_3}{9}+\left(i+ic_1^2+i\left(c_1+c_2\right)\left(c_2-1\right)-\frac{25ia_2}{36}+\frac{70ia_2^2}{27}\right.\\
	&\left.+\frac{185ia_3}{36}\right)mq+\left(\frac{7}{4}-\frac{25m^2}{18}+\frac{a_2}{2}+\frac{a_2 \mathcal{H}_{(2m)}}{3}+\frac{55a_2 m^2}{54}\right)q^2.
\end{align}

\section{Expressions of the solutions \texorpdfstring{ $R^{(in)}_{2m\omega}$}{} and \texorpdfstring{ $X^{(in)}_{2m\omega}$}{}}\label{ApenC}
The real parts of $\xi_{\ell m}$ are as follows
{\small{
		\begin{align}
			\nonumber f_{2m}^{(2)} = & \left( \frac{383 a_{2}^2}{5670}-\frac{65 a_{2}}{42}+\frac{463 a_{3}}{3024}-\frac{149 m^2 q^2}{5670}-\frac{4 q^2}{63}+\frac{911}{252} \right) j_{0} +  \Big( \frac{925 a_{2}}{882} -\frac{701 a_{2}^2}{19845} - \frac{1165 a_{3}}{10584} \\
			\nonumber & -\frac{3733 m^2 q^2}{79380}+\frac{851 q^2}{17640}-\frac{16949}{8820} \Big) j_{2} + \bigg( \frac{1}{49} -\frac{17 a_{2}^2}{13230}-\frac{10 a_{2}}{441}-\frac{59 a_{3}}{2352}-\frac{71 m^2 q^2}{8820}-\frac{q^2}{392} \bigg) j_{4} \\
			\nonumber &  + \bigg( \frac{a_{2}}{21} - \frac{16 a_{2}^2}{945} -\frac{5 a_{3}}{126}-\frac{2 m^2 q^2}{945}+\frac{2 q^2}{105}-\frac{107}{210} \bigg) j_{2} ln z + \bigg(\frac{32 a_{2}^2}{315}-\frac{55 a_{2}}{21}+\frac{5 a_{3}}{21}+\frac{4 m^2 q^2}{315}\\
			\nonumber&- \frac{4 q^2}{35} + \frac{389}{70}\bigg) n_{1} + \left(\frac{32 a_{2}^2}{945}-\frac{2 a_{2}}{21}+\frac{5 a_{3}}{63}+\frac{4 m^2 q^2}{945}-\frac{4 q^2}{105}+\frac{107}{105}\right) \mathfrak{D}_{-3}^{nj} + \left(\frac{a_{2} m q}{81}+\frac{7 m q}{180}\right) j_{3}\\
			\nonumber & + 6 \mathfrak{D}_{-1}^{nj} + \left(9-\frac{2 a_{2}}{15}\right) \mathfrak{D}_{1}^{nj} + \left(\frac{14 a_{2}}{45}-\frac{1}{3}\right) \mathfrak{D}_{3}^{nj} - \frac{m q}{10}  j_{1}-\frac{m q}{5}  j_{1} ln z -\frac{13 m q}{90} j_{3} ln z-\frac{1}{2} j_{2} (ln z)^{2} \\
			& +4 \mathfrak{D}_{2}^{nnj} , \\
			\nonumber f_{3m}^{(2)} = & \left(\frac{401 a_{2}^2}{1890}-\frac{9473 a_{2}}{1260}+\frac{53 a_{3}}{112}-\frac{m^2 q^2}{60}+\frac{3161}{140}\right) j_{1} + \left( \frac{4819 a_{2}}{3780} - \frac{571 a_{2}^2}{17010}-\frac{877 a_{3}}{9072}-\frac{349 m^2 q^2}{45360}  \right. \\
			\nonumber & \left. -\frac{q^2}{360} - \frac{269}{140} \right) j_{3} + \left(\frac{2 a_{2}^2}{1701}-\frac{199 a_{2}}{7560}-\frac{65 a_{3}}{4536}-\frac{17 m^2 q^2}{5040}-\frac{q^2}{360}+\frac{1}{36}\right) j_{5} + \left(\frac{a_{2}}{42} - \frac{5 a_{2}^2}{378} - \frac{5 a_{3}}{168}  \right.  \\
			\nonumber & \left. +\frac{m^2 q^2}{1512} - \frac{13}{42} \right) j_{3} ln z + \left( \frac{2015 a_{2}}{504} - \frac{65 a_{2}^2}{567} - \frac{65 a_{3}}{252}+\frac{13 m^2 q^2}{2268}-\frac{3511}{252}\right) n_{0} + \left( \frac{25 a_{2}^2}{189} - \frac{135 a_{2}}{28} \right. \\
			\nonumber & \left. +\frac{25 a_{3}}{84} - \frac{5 m^2 q^2}{756} + \frac{445}{42} \right) n_{2} + \left( \frac{a_{2}}{21} - \frac{5 a_{2}^2}{189} -\frac{5 a_{3}}{84}+\frac{m^2 q^2}{756}-\frac{13}{21} \right) \mathfrak{D}_{-4}^{nj} - \left( \frac{11 a_{2} m q}{2268} + \frac{m q}{36} \right) j_{2}  \\
			\nonumber & +\left(\frac{a_{2} m q}{168}+\frac{3 m q}{160}\right) j_{4} + \left(\frac{355}{21}-\frac{8 a_{2}}{63}\right) \mathfrak{D}_{2}^{nj}+\left(\frac{11 a_{2}}{42}-\frac{3}{7}\right) \mathfrak{D}_{4}^{nj} - \frac{13}{126} m q j_{2} ln z - \frac{5}{56} m q j_{4} ln z  \\
			& -\frac{1}{2} j_{3} (ln z)^{2} - 30 \mathfrak{D}_{-2}^{nj}-17 \mathfrak{D}_{0}^{nj}+4 \mathfrak{D}_{3}^{nnj} ,
		\end{align}
}}
{\small{
		\begin{align}
			\nonumber f_{2m}^{(3)} = & \left(\frac{9}{4}-\frac{a_{2}}{30}\right) j_{1} (ln z)^{2} + \left(\frac{7 a_{2}}{90}-\frac{1}{12}\right) j_{3} (ln z)^{2} - \frac{3}{2} n_{0} (ln z)^{2} + \mathfrak{D}_{2}^{nj} (ln z)^{2} - \left(\frac{a_{2} m q}{60} + \frac{191 m q}{180}\right) j_{0} ln z \\
			\nonumber & +\left(\frac{5 a_{2}^2}{63}-\frac{16 a_{2}^3}{14175}+\frac{2 q^2 a_{2}}{1575}-\frac{a_{3} a_{2}}{378}-\frac{887 a_{2}}{3150}-\frac{2 m^2 q^2 a_{2}}{14175}+\frac{m^2 q^2}{105}+\frac{5 a_{3}}{28}+\frac{\mathbb{T}_{1}}{5}-\frac{q^2}{28}+\frac{321}{140}\right)  \\
			\nonumber & \times j_{1} ln z+\left(\frac{m q a_{2}^2}{135} +\frac{m q a_{2}}{54}+\frac{17 a_{3} m q}{1512}+\frac{293 m q}{252}\right) j_{2} ln z+\left(\frac{16 a_{2}^3}{6075}-\frac{11 a_{2}^2}{315}+\frac{2 m^2 q^2 a_{2}}{6075}+\frac{a_{3} a_{2}}{162} \right. \\
			\nonumber & \left.-\frac{2 q^2 a_{2}}{675} +\frac{1543 a_{2}}{18900}+\frac{169 q^2}{2520}+\frac{\mathbb{T}_{1}}{5}-\frac{10 a_{3}}{189}-\frac{107}{1260}-\frac{13 m^2 q^2}{3780}\right) j_{3} ln z + \left(\frac{43 a_{2} m q}{3780}-\frac{11 m q}{420}\right)\\
			\nonumber & \times j_{4} ln z + \left(\frac{a_{2}}{7} -\frac{16 a_{2}^2}{315}+\frac{2 q^2}{35}-\frac{5 a_{3}}{42}-\frac{107}{70}-\frac{2 m^2 q^2}{315}\right) n_{0} ln z - \frac{13 m q}{6} n_{1} ln z + \frac{2 m q}{5} \mathfrak{D}_{1}^{nj} ln z  \\ 
			\nonumber & + \left( \frac{32 a_{2}^2}{945} - \frac{2 a_{2}}{21} +\frac{4 m^2 q^2}{945}+\frac{5 a_{3}}{63}-\frac{4 q^2}{105}+\frac{107}{105}\right) \mathfrak{D}_{2}^{nj} ln z + \frac{13 m q}{45} \mathfrak{D}_{3}^{nj} ln z + \left(\frac{m q}{8}  -\frac{26 m q a_{2}^2}{1215}\right. \\ 
			\nonumber & \left.+\frac{391 m q a_{2}}{22680}+\frac{m q \mathbb{T}_{1}}{15}  -\frac{71 a_{3} m q}{2268}\right) j_{0} + \left(\frac{21}{100} -\frac{2381 a_{2}^3}{66150}+\frac{8207 a_{2}^2}{132300}+\frac{1363 q^2 a_{2}}{29400}-\frac{185 a_{3} a_{2}}{2352}\right. \\ 
			\nonumber & \left.-\frac{83821 a_{2}}{44100}-\frac{1201 m^2 q^2 a_{2}}{88200}-\frac{187 a_{3}}{168}-\frac{2153 q^2}{8400}-\frac{10681 m^2 q^2}{37800}\right) j_{1} + \left(\frac{13 m q a_{2}^2}{630} -\frac{223 m q a_{2}}{7938} \right. \\
			\nonumber & \left. +\frac{2089 a_{3} m q}{63504}+\frac{17 m q}{336}+\frac{11 m q \mathbb{T}_{1}}{126}\right) j_{2}+\left(\frac{97 a_{2}^3}{36450}+\frac{5339 a_{2}^2}{170100}-\frac{\mathcal{H}_{(2m)}}{648}  q^2 a_{2}+\frac{35 a_{3} a_{2}}{3888}-\frac{19 q^2 a_{2}}{6300}\right. \\
			\nonumber & \left.+\frac{9053 a_{2}}{25200}-\frac{673 m^2 q^2 a_{2}}{1020600}+\frac{53507 m^2 q^2}{680400} +\frac{7891 q^2}{151200}+\frac{4609 a_{3}}{18144}-\frac{457}{1050}\right) j_{3} + \left(\frac{m q}{112}-\frac{41 m q a_{2}^2}{17010}\right. \\
			\nonumber & \left.-\frac{53 m q a_{2}}{17640}+\frac{13 m q \mathbb{T}_{1}}{630}-\frac{23 a_{3} m q}{7056}\right) j_{4} +\left(\frac{139 a_{2}^2}{476280}-\frac{11 a_{2}^3}{142884}+\frac{143 m^2 q^2 a_{2}}{381024}+\frac{\mathcal{H}_{(2m)} q^2 a_{2}}{12960} \right. \\
			\nonumber & \left.+\frac{53 q^2 a_{2}}{79380}  -\frac{43 a_{3} a_{2}}{95256}-\frac{277 a_{2}}{105840}+\frac{1}{504}-\frac{m^2 q^2}{648} - \frac{q^2}{1680}-\frac{109 a_{3}}{45360}\right) j_{5} + \left(\frac{193 a_{2}^3}{10206}-\frac{110 a_{2}^2}{1701}\right. \\
			\nonumber & \left.+\frac{361 m^2 q^2 a_{2}}{20412}   +\frac{\mathcal{H}_{(2m)} q^2 a_{2}}{540} +\frac{1033 a_{3} a_{2}}{27216}+\frac{6911 a_{2}}{5670}- \frac{733 q^2 a_{2}}{22680} +\frac{6119 m^2 q^2}{68040}+\frac{2441 q^2}{15120}\right. \\
			\nonumber & \left.+\frac{48353 a_{3}}{90720}-\frac{2539}{3780}\right) n_{0}+ \left(\frac{m q}{12}-\frac{2 m q a_{2}^2}{45} +\frac{8 m q a_{2}}{189}- \frac{17 a_{3} m q}{252}\right) n_{1}+ \left( \frac{65 a_{2}^2}{1134} \right.  \left. - \frac{32 a_{2}^3}{1215} + \frac{4 q^2 a_{2}}{135}\right. \\
			\nonumber & \left.-\frac{5 a_{3} a_{2}}{81}-\frac{1574 a_{2}}{945}-\frac{4 m^2 q^2 a_{2}}{1215}+\frac{197}{126}-\frac{61 q^2}{504}-\frac{631 m^2 q^2}{2268}-\frac{2455 a_{3}}{3024}\right) n_{2} + \left(\frac{32 a_{2}^3}{6075}-\frac{58 a_{2}^2}{2835}\right. \\
			\nonumber & \left.+\frac{4 m^2 q^2 a_{2}}{6075}+\frac{a_{3} a_{2}}{81}  -\frac{4 q^2 a_{2}}{675}+\frac{824 a_{2}}{4725}+\frac{2 q^2}{315}-\frac{5 a_{3}}{378}-\frac{107}{630}-\frac{2 m^2 q^2}{2835}\right) \mathfrak{D}_{-4}^{nj} -12 \mathfrak{D}_{-1}^{nnj} -\left(\frac{2 m q a_{2}^2}{135}\right. \\
			\nonumber & \left. + \frac{m q a_{2}}{21}+ \frac{m q}{6}  + \frac{17 a_{3} m q}{756}\right) \mathfrak{D}_{-3}^{nj} + \left(\frac{4 q^2 a_{2}}{1575} - \frac{32 a_{2}^3}{14175}-\frac{2 a_{2}^2}{45} -\frac{a_{3} a_{2}}{189}+\frac{7468 a_{2}}{1575}-\frac{4 m^2 q^2 a_{2}}{14175}+ \frac{2 q^2}{35} \right. \\
			\nonumber & \left.-\frac{5 a_{3}}{42}-\frac{457}{70} -\frac{2 m^2 q^2}{315} \right) \mathfrak{D}_{-2}^{nj} + \left(\frac{59 a_{2}}{21} -\frac{19 a_{2}^2}{567}+\frac{37 m^2 q^2}{567}+\frac{4 q^2}{315}-\frac{2629}{630}-\frac{103 a_{3}}{1512}\right) \mathfrak{D}_{0}^{nj} - 8 \mathfrak{D}_{2}^{nnnj} \\
			\nonumber & + \frac{1}{5} m q \mathfrak{D}_{1}^{nj}+\left(\frac{1402 a_{2}^2}{19845}-\frac{925 a_{2}}{441}+\frac{3733 m^2 q^2}{39690}+\frac{1165 a_{3}}{5292}+\frac{16949}{4410}-\frac{851 q^2}{8820}\right) \mathfrak{D}_{2}^{nj} - \left(\frac{2 a_{2} m q}{81}\right. \\
			\nonumber & \left. +  \frac{7 m q}{90}\right) \mathfrak{D}_{3}^{nj}+\left(\frac{17 a_{2}^2}{6615}+\frac{20 a_{2}}{441}+\frac{71 m^2 q^2}{4410}+\frac{q^2}{196}+\frac{59 a_{3}}{1176}-\frac{2}{49}\right) \mathfrak{D}_{4}^{nj} + \left(\frac{4 a_{2}}{21}-\frac{64 a_{2}^2}{945}+\frac{8 q^2}{105}\right. \\
			\nonumber & \left.-\frac{10 a_{3}}{63}  -\frac{214}{105}-\frac{8 m^2 q^2}{945}\right) \mathfrak{D}_{2}^{njj} + \left(\frac{4 a_{2}}{21}-\frac{64 a_{2}^2}{945}+\frac{8 q^2}{105}-\frac{10 a_{3}}{63}-\frac{214}{105}-\frac{8 m^2 q^2}{945}\right) \mathfrak{D}_{-3}^{nnj}
			\\&+ \left(\frac{4 a_{2}}{15}-18\right) \mathfrak{D}_{1}^{nnj}+\left(\frac{2}{3}-\frac{28 a_{2}}{45}\right) \mathfrak{D}_{3}^{nnj} ,
		\end{align}
}}
and	the imaginary parts are
\begin{align}
	g_{\ell m}^{(2)} = & f_{\ell m}^{(1)} ln z - \frac{\mathbb{T}_{1}}{z} j_{\ell} + \varsigma_{\ell m}^{(2)} , \\
	g_{\ell m}^{(3)} = & f_{\ell m}^{(2)} ln z - \frac{\mathbb{T}_{1} f_{\ell m}^{(1)}}{z}  -  \frac{\mathbb{T}_{2} j_{\ell}}{2 z^{2}} + \frac{1}{3} j_{\ell} (ln z)^{3} + \varsigma_{\ell m}^{(3)} ,
\end{align}
where
{\small{
		\begin{align}
			\nonumber \varsigma_{2 m}^{(2)} = & \left( \frac{2 a_{2}^2}{81} + \frac{5 a_{3}}{108} + \frac{a_{2}}{180} + \frac{m^2 q^2}{324}-\frac{23 q^2}{360} \right) j_{3} + \frac{a_{2}}{30} j_{1} + \left(\frac{a_{2} m q}{60}+\frac{191 m q}{180}\right) j_{0} + \left(\frac{a_{2} m q}{189}-\frac{68 m q}{63}\right)j_{2}  \\
			& +\left(\frac{11 m q}{420}-\frac{43 a_{2} m q}{3780}\right) j_{4} - \frac{q^2 }{20} j_{1} + \frac{13 m q }{6} n_{1}-\frac{2 m q}{5}  \mathfrak{D}_{1}^{nj}-\frac{13}{45} m q \mathfrak{D}_{3}^{nj} , \\
			\nonumber \varsigma_{2 m}^{(3)} = & \left( \frac{a_{2}}{72} -\frac{44 a_{2}^3}{1701}+\frac{2083 a_{2}^2}{11340}+\frac{1007 q^2 a_{2}}{52920}-\frac{50 a_{3} a_{2}}{567}-\frac{17 \mathcal{H}_{(2m)} q^2 a_{2}}{3780}-\frac{41 m^2 q^2 a_{2}}{3969}+\frac{31 m^2 q^2}{3240}+\frac{263 a_{3}}{756}
			\right. \\
			\nonumber & \left. -\frac{367 q^2}{720}\right) j_{0} + \left(\frac{4136 m q a_{2}^2}{99225}-\frac{39671 m q a_{2}}{52920}+\frac{3863 a_{3} m q}{42336}+\frac{257603 m q}{88200}+\frac{m q \mathbb{T}_{1}}{70}\right) j_{1} + \left(\frac{419 a_{2}^3}{19845} \right. \\
			\nonumber & \left. -\frac{5017 a_{2}^2}{39690}+\frac{100 a_{3} a_{2}}{1323}-\frac{23 a_{2}}{336}-\frac{95 q^2 a_{2}}{12348}-\frac{13 \mathcal{H}_{(2m)} q^2 a_{2}}{52920}-\frac{18311 m^2 q^2 a_{2}}{3333960}+\frac{881 q^2}{2016}-\frac{307 a_{3}}{1323}  \right. \\
			\nonumber & \left. -\frac{293 m^2 q^2}{9072}\right) j_{2}+\left(-\frac{2087 m q a_{2}^2}{510300}+\frac{3659 m q a_{2}}{34020}+\frac{m q \mathbb{T}_{1}}{270}-\frac{31 a_{3} m q}{2268}-\frac{28171 m q}{75600}\right) j_{3} + \left(\frac{5 a_{2}^3}{11907}  \right. \\
			\nonumber & \left. +\frac{83 a_{2}^2}{26460}+\frac{353 m^2 q^2 a_{2}}{666792}+\frac{2701 q^2 a_{2}}{370440}+\frac{20 a_{3} a_{2}}{3969}+\frac{a_{2}}{1008}-\frac{\mathcal{H}_{(2m)} q^2 a_{2}}{1960}+\frac{67 m^2 q^2}{15120}+\frac{13 a_{3}}{1764}-\frac{13 q^2}{1120}\right)  \\
			\nonumber & \times j_{4}- \left( \frac{m q a_{2}^2}{1428840} + \frac{179 m q a_{2}}{79380} - \frac{181 m q}{70560} - \frac{m q \mathbb{T}_{1}}{1890} + \frac{955 a_{3} m q}{762048} \right) j_{5} + \left(\frac{4177 m q a_{2}}{11340} - \frac{1703 m q a_{2}^2}{102060}  \right. \\
			\nonumber & \left. -\frac{m q \mathbb{T}_{1}}{90}-\frac{565 a_{3} m q}{13608}-\frac{33847 m q}{22680}\right) n_{0} + \left( \frac{296 a_{2}^2}{945} - \frac{44 a_{2}^3}{945} - \frac{\mathcal{H}_{(2m)} q^2 a_{2}}{210} + \frac{5 q^2 a_{2}}{147} + \frac{a_{2}}{12} -\frac{10 a_{3} a_{2}}{63}  \right. \\ 
			\nonumber & \left.-  \frac{19 m^2 q^2 a_{2}}{4410}+\frac{5 m^2 q^2}{108}+\frac{37 a_{3}}{63}-\frac{23 q^2}{24}\right) n_{1} +\left(\frac{208 m q a_{2}^2}{8505}-\frac{353 m q a_{2}}{756}+\frac{65 a_{3} m q}{1134}+\frac{6247 m q}{3780}\right) n_{2} \\ 
			\nonumber & + \left(\frac{13 m q a_{2}}{945} -\frac{208 m q a_{2}^2}{42525}  -\frac{13 a_{3} m q}{1134}-\frac{1391 m q}{9450}\right) \mathfrak{D}_{-4}^{nj} + \left(\frac{5 q^2 a_{2}}{441} - \frac{44 a_{2}^3}{2835}-\frac{2 a_{2}^2}{105}-\frac{ \mathcal{H}_{(2m)} q^2 a_{2}}{630} \right. \\
			\nonumber & \left.  - \frac{10 a_{3} a_{2}}{189}-\frac{19 m^2 q^2 a_{2}}{13230}-\frac{a_{3}}{28}\right) \mathfrak{D}_{-3}^{nj} + \left(\frac{2 m q a_{2}}{105}-\frac{32 m q a_{2}^2}{4725}-\frac{a_{3} m q}{63}-\frac{794 m q}{175}\right) \mathfrak{D}_{-2}^{nj} - \left(\frac{a_{2} m q}{30}\right. \\
			\nonumber &\left. +\frac{191 m q}{90}\right) \mathfrak{D}_{0}^{nj}  +\left(\frac{q^2}{10}-\frac{a_{2}}{15}\right) \mathfrak{D}_{1}^{nj}+\left(\frac{136 m q}{63}-\frac{2 a_{2} m q}{189}\right) \mathfrak{D}_{2}^{nj} + \left(\frac{23 q^2}{180}-\frac{4 a_{2}^2}{81}-\frac{a_{2}}{90}-\frac{5 a_{3}}{54}\right. \\
			\nonumber & \left.-\frac{m^2 q^2}{162}\right) \mathfrak{D}_{3}^{nj}+\left(\frac{43 a_{2} m q}{1890}-\frac{11 m q}{210}\right) \mathfrak{D}_{4}^{nj}+\frac{4}{5} m q \mathfrak{D}_{1}^{nnj}+\frac{26}{45} m q \mathfrak{D}_{3}^{nnj} + \Bigg[ \left(\frac{a_{2} m q}{105}-\frac{16 a_{2}^2 m q}{4725}  \right.\\
			\nonumber &\left. -\frac{a_{3} m q}{126}  -\frac{107 m q}{1050}\right) j_{1} + \left(\frac{13 a_{2} m q}{1890}-\frac{104 a_{2}^2 m q}{42525} - \frac{13 a_{3} m q}{2268}-\frac{1391 m q}{18900}\right) j_{3} + \bigg(\frac{22 a_{2}^3}{2835}+\frac{a_{2}^2}{105}\\
			& +\frac{5 a_{2} a_{3}}{189}+\frac{a_{2} \mathcal{H}_{(2m)} q^2}{1260}+\frac{19 a_{2} m^2 q^2}{26460}-\frac{5 a_{2} q^2}{882}+\frac{a_{3}}{56}\bigg) j_{2} \Bigg] ln z + \left( \frac{m q}{10} j_{1}+\frac{13 m q}{180} j_{3} \right) (ln z)^{2} ,
		\end{align}
}}
with	
\begin{align}	
	\mathbb{T}_{1} = & 1 + c_{1}^{2} - (c_{1} + c_{2})(1 - c_{2}), \\ \mathbb{T}_{2} = & c_{1} c_{2} (-c_{1}-c_{2}+3)+2 (c_{1}-1) c_{1}+2 (c_{2}-1) c_{2} + 1 .
\end{align}

Inserting above results into the assumed form of the solution $X^{in}_{\ell m \omega}$, and expanding it in terms of $\eta$, we acquire $X^{in}_{2 m \omega}$ as
{\small{
		\begin{align}
			\nonumber
			X^{in}_{2m\omega}=&\frac{z^3}{15}-\frac{z^5}{210}+\frac{z^7}{7560}-\frac{z^9}{498960}+\frac{z^{11}}{51891840}-\frac{z^{13}}{7783776000}+\frac{z^{15}}{1587890304000}\\
			\nonumber &+\eta\left[\left(\frac{a_2}{45}+\frac{imq}{30}\right)z^2+\left(-\frac{13}{630}-\frac{a_2}{270}-\frac{11imq}{3780}\right)z^4+\left(\frac{1}{810}+\frac{11a_2}{68040}+\frac{13imq}{136080}\right)z^6+\left(-\frac{53}{1782000}\right.\right.\\
			\nonumber&\left.\left.-\frac{a_2}{299376}-\frac{imq}{598752}\right)z^8+\left(\frac{227}{567567000}+\frac{19a_2}{467026560}+\frac{17imq}{934053120}\right)z^{10}+\left(-\frac{11911}{3432645216000}\right.\right.\\
			\nonumber&\left.\left.-\frac{23a_2}{70053984000}-\frac{19imq}{140107968000}\right)z^{12}\right]+\eta^2\left[\left(\frac{a_2^2}{90}+\frac{a_3}{48}+\frac{imq}{60}+\frac{ia_2mq}{180}+\frac{q^2}{120}-\frac{m^2 q^2}{120}\right)z\right.\\
			\nonumber&\left.+\left(\frac{ia_2}{90}-\frac{mq}{30}\right)z^2+\left(\frac{26743}{110250}-\frac{4546a_2}{99225}+\frac{13483a_2^2}{5953500}+\frac{2767a_3}{635040}-\frac{433imq}{22680}-\frac{13ia_2mq}{22680}-\frac{12329q^2}{2646000}\right.\right.\\
			\nonumber&\left.\left.+\frac{28079m^2q^2}{23814000}-\frac{107lnz}{3150}+\frac{a_2lnz}{315}-\frac{16a_2^2lnz}{14175}-\frac{a_3lnz}{378}+\frac{2q^2lnz}{1575}-\frac{2m^2q^2lnz}{14175}\right)z^3+\left(-\frac{ia_2}{945}+\frac{2ia_2^2}{8505}\right.\right.\\
			\nonumber&\left.\left.+\frac{ia_3}{2268}+\frac{mq}{270}+\frac{a_2mq}{8505}-\frac{iq^2}{7560}+\frac{im^2q^2}{34020}\right)z^4+\left(-\frac{140953}{9261000}+\frac{31291a_2}{8334900}-\frac{112943a_2^2}{500094000}-\frac{307a_3}{658560}\right.\right.\\
			\nonumber&\left.\left.+\frac{17imq}{12960}+\frac{17ia_2mq}{816480}+\frac{2717q^2}{8232000}-\frac{13051m^2q^2}{222264000}+\frac{107lnz}{44100}-\frac{a_2lnz}{4410}+\frac{8a_2^2lnz}{99225}+\frac{a_3lnz}{5292}-\frac{q^2lnz}{11025}\right.\right.\\
			\nonumber&\left.\left.+\frac{m^2q^2lnz}{99225}\right)z^5+\left(\frac{ia_2}{27216}-\frac{ia_2^2}{76545}-\frac{ia_3}{40824}-\frac{19mq}{136080}-\frac{a_2mq}{153090}+\frac{iq^2}{136080}-\frac{im^2q^2}{612360}\right)z^6\right.\\
			\nonumber&\left.+\left(\frac{255779}{687629250}-\frac{314581a_2}{2750517000}+\frac{709229a_2^2 }{99018612000}+\frac{161321a_3}{10561985280}-\frac{1087imq}{49896000}-\frac{ia_2mq}{2566080}\right.\right.\\
			\nonumber&\left.\left.-\frac{416827q^2 }{44008272000}+\frac{571177 m^2q^2}{396074448000}-\frac{107lnz}{1587600}+\frac{a_2lnz}{158760}-\frac{2a_2^2lnz}{893025}-\frac{a_3lnz}{190512}+\frac{q^2lnz}{396900}\right.\right.\\
			\nonumber&\left.\left.-\frac{m^2q^2lnz}{3572100}\right)z^7+\left(-\frac{ia_2}{1496880}+\frac{ia_2^2}{3367980}+\frac{ia_3}{1796256}+\frac{mq}{374220}+\frac{a_2mq}{6735960}-\frac{iq^2}{5987520}\right.\right.\\
			\nonumber&\left.\left.+\frac{im^2q^2}{26943840}\right)z^8+\left(-\frac{46739723}{9439774344000}+\frac{2918507a_2}{1573295724000}-\frac{40008173a_2^2}{339831876384000}\right.\right.\\
			\nonumber&\left.\left.-\frac{9280457 a_3}{36248733480960}+\frac{539597imq}{980755776000}+\frac{5ia_2mq}{1120863744}+\frac{22438249q^2}{151036389504000}\right.\right.\\
			\nonumber&\left.\left.-\frac{28259449m^2q^2}{1359327505536000}+\frac{107lnz}{104781600}-\frac{a_2lnz}{10478160}+\frac{a_2^2lnz}{29469825}+\frac{a_3lnz}{12573792}-\frac{q^2lnz}{26195400}\right.\right.\\
			\nonumber&\left.\left.+\frac{m^2q^2lnz}{235758600}\right)z^9\right]+\eta^3\left[-\frac{a_2}{45}+\frac{a_2^2}{1620}+\frac{a_2^3}{162}+\frac{a_3}{135}+\frac{a_2a_3}{48}-\frac{imq}{90}-\frac{ia_2mq}{216}-\frac{4ia_2^2mq}{1215}\right.\\
			\nonumber&\left.-\frac{7ia_3mq}{2592}-\frac{q^2}{30}+\frac{13a_2q^2}{1890}-\frac{a_2\mathcal{H}_{(2m)}q^2}{540}+\frac{m^2q^2}{40}-\frac{17a_2m^2q^2}{1890}+\left(\frac{ia_2^2}{180}+\frac{ia_3}{72}-\frac{mq}{36}-\frac{a_2mq}{360}\right.\right.\\
			\nonumber&\left.\left.+\frac{4a_2^2mq}{1215}+\frac{a_3mq}{162}-\frac{iq^2}{120}+\frac{43ia_2q^2}{7560}-\frac{ia_2\mathcal{H}_{(2m)}q^2}{540}-\frac{ia_2m^2q^2}{126}\right)z+\left(\frac{319}{6300}+\frac{41551a_2}{661500}-\frac{7597a_2^2}{476280}\right.\right.\\
			\nonumber&\left.\left.+\frac{6553a_2^3}{17860500}-\frac{131a_3}{15120}-\frac{803a_2a_3}{1905120}+\frac{2074imq}{18375}-\frac{75469ia_2mq}{4762800}+\frac{24889ia_2^2mq}{5953500}+\frac{2929ia_3mq}{423360}\right.\right.\\
			\nonumber&\left.\left.-\frac{79q^2}{8400}-\frac{2204a_2q^2}{496125}+\frac{a_2\mathcal{H}_{(2m)}q^2}{1080}+\frac{191m^2q^2}{113400}+\frac{139957a_2m^2q^2}{35721000}-\frac{107a_2lnz}{9450}+\frac{a_2^2lnz}{945}-\frac{16a_2^3lnz}{42525}\right.\right.\\
			\nonumber&\left.\left.-\frac{a_2a_3lnz}{1134}-\frac{107imqlnz}{6300}+\frac{ia_2mqlnz}{630}-\frac{8ia_2^2mqlnz}{14175}-\frac{ia_3mqlnz}{756}+\frac{2a_2q^2lnz}{4725}-\frac{2a_2m^2q^2lnz}{42525}\right)z^2\right.\\
			\nonumber&\left.+\left(-\frac{319ia_2}{45360}+\frac{25873ia_2^2}{35721000}-\frac{6959ia_2^3}{4465125}+\frac{5287ia_3}{4762800}-\frac{302ia_2a_3}{59535}+\frac{3271mq}{226800}-\frac{46507a_2mq}{14288400}\right.\right.\\
			\nonumber&\left.\left.-\frac{2407a_2^2mq}{1275750}-\frac{13873a_3mq}{4762800}-\frac{13iq^2}{12960}+\frac{523ia_2q^2}{740880}-\frac{11ia_2\mathcal{H}_{(2m)}q^2}{992250}+\frac{199im^2q^2}{408240}\right.\right.\\
			\nonumber&\left.\left.+\frac{155867ia_2m^2q^2}{250047000}+\frac{ia_2^2lnz}{1575}+\frac{22ia_2^3lnz}{42525}+\frac{ia_3lnz}{840}+\frac{ia_2a_3lnz}{567}+\frac{mqlnz}{180}+\frac{a_2mqlnz}{630}+\frac{a_2^2mqlnz}{2025}\right.\right.\\
			\nonumber&\left.\left.+\frac{17a_3mqlnz}{22680}-\frac{ia_2q^2lnz}{2646}+\frac{ia_2\mathcal{H}_{(2m)}q^2lnz}{18900}+\frac{19ia_2m^2q^2lnz}{396900}\right)z^3+\left(-\frac{292088}{3472875}+\frac{170483a_2 }{166698000}\right.\right.\\
			\nonumber&\left.\left.+\frac{12749a_2^2}{12348000}-\frac{7313a_2^3}{30618000}-\frac{52559a_3 }{26671680}-\frac{499a_2a_3}{1088640}-\frac{1118863imq}{166698000}+\frac{605681ia_2mq}{400075200}\right.\right.\\
			\nonumber&\left.\left.-\frac{30991ia_2^2mq}{83349000}-\frac{200519ia_3mq }{320060160}+\frac{145351q^2}{55566000}+\frac{13177a_2q^2 }{23814000}-\frac{5a_2\mathcal{H}_{(2m)}q^2}{54432}-\frac{883979m^2q^2}{2000376000}\right.\right.\\
			\nonumber&\left.\left.-\frac{4597a_2m^2q^2}{13395375}+\frac{1391lnz}{132300}+\frac{359a_2lnz}{396900}+\frac{103a_2^2lnz}{595350}+\frac{8a_2^3lnz}{127575}+\frac{13a_3lnz}{15876}+\frac{a_2a_3lnz}{6804}\right.\right.\\
			\nonumber&\left.\left.+\frac{1177imqlnz}{793800}-\frac{11ia_2mqlnz}{79380}+\frac{44ia_2^2mqlnz}{893025}+\frac{11ia_3mqlnz}{95256}-\frac{13q^2lnz}{33075}-\frac{a_2q^2lnz}{14175}+\frac{13m^2q^2lnz}{297675}\right.\right.\\
			\nonumber&\left.\left.+\frac{a_2m^2q^2lnz}{127575}\right)z^4+\left(\frac{143ia_2}{272160}-\frac{12679ia_2^2}{111132000}+\frac{68821ia_2^3}{562605750}-\frac{9203ia_3}{44452800}+\frac{1514ia_2a_3}{3750705}-\frac{799mq}{793800}\right.\right.\\
			\nonumber&\left.\left.+\frac{97079a_2mq}{400075200}+\frac{21283a_2^2mq}{160744500}+\frac{40279a_3mq}{200037600}+\frac{iq^2}{7560}-\frac{42911ia_2q^2}{560105280}+\frac{883ia_2\mathcal{H}_{(2m)}q^2}{111132000}-\frac{11im^2q^2}{272160}\right.\right.\\
			\nonumber&\left.\left.-\frac{122687ia_2m^2q^2}{7876480500}-\frac{ia_2^2lnz}{22050}-\frac{11ia_2^3lnz}{297675}-\frac{ia_3lnz}{11760}-\frac{ia_2a_3lnz}{7938}-\frac{mqlnz}{2520}-\frac{a_2mqlnz}{8820}-\frac{a_2^2mqlnz}{28350}\right.\right.\\
			\nonumber&\left.\left.-\frac{17a_3mqlnz}{317520}+\frac{ia_2q^2lnz}{37044}-\frac{ia_2\mathcal{H}_{(2m)}q^2lnz}{264600}-\frac{19ia_2m^2q^2lnz}{5556600}\right)z^5+\left(\frac{11392303}{2357586000}-\frac{7019189a_2}{24754653000}\right.\right.\\
			\nonumber&\left.\left.-\frac{9777269a_2^2}{594111672000}+\frac{1014359 a_2^3}{81015228000}+\frac{432581a_3}{2829103200}+\frac{2453041a_2a_3}{95057867520}+\frac{13269859imq}{99018612000}\right.\right.\\
			\nonumber&\left.\left.-\frac{20745451 ia_2mq}{396074448000}+\frac{11251921ia_2^2mq}{891167508000}+\frac{4061513 ia_3mq}{190115735040}-\frac{2560147q^2}{18860688000}-\frac{282091 a_2q^2}{12377326500}\right.\right.\\
			\nonumber&\left.\left.+\frac{61a_2\mathcal{H}_{(2m)}q^2}{17962560}+\frac{160477 m^2q^2}{7715736000}+\frac{1934771a_2m^2q^2}{162030456000}-\frac{107lnz}{170100}-\frac{337a_2lnz}{14288400}-\frac{283a_2^2lnz}{21432600}\right.\right.\\
			\nonumber&\left.\left.-\frac{22a_2^3lnz}{8037225}-\frac{a_3lnz}{20412}-\frac{11a_2a_3lnz}{1714608}-\frac{1391imqlnz}{28576800}+\frac{13ia_2mqlnz}{2857680}-\frac{13ia_2^2mqlnz}{8037225}-\frac{13ia_3mqlnz}{3429216}\right.\right.\\
			&\left.\left.+\frac{q^2lnz}{42525}+\frac{11a_2q^2lnz}{3572100}-\frac{m^2q^2lnz}{382725}-\frac{11a_2m^2q^2lnz}{32148900}\right)z^6\right].
		\end{align}
}}

Furthermore, using the transformation between the two types of equations, we can obtain
the corresponding solutions $R^{(in)}_{2 m \omega}$ of the Teukolsky-like equation without the source, which are given by
{\small{\
		\begin{align}\label{R1eq}
			\nonumber
			\omega R^{in}_{2m\omega}=&\frac{z^4}{30} + \frac{i z^5}{45} - \frac{11 z^6}{1260} - \frac{i z^7}{420} + \frac{23 z^8}{45360} + \frac{i z^9}{11340} - \frac{13 z^{10}}{997920} - \frac{i z^{11}}{598752} + \frac{59 z^{12}}{311351040}\\
			\nonumber&+\frac{i z^{13}}{51891840}-\frac{83z^{14}}{46702656000}-\frac{iz^{15}}{6671808000}+\eta\left[\left( -\frac{1}{15} + \frac{a_{2}}{180}- \frac{i m q}{60} \right)z^3  +\left( -\frac{i}{60} + \frac{i a_{2}}{180}\right.\right.\\
			\nonumber&\left.\left.  + \frac{m q}{45} \right) z^4  +  \left( -\frac{41}{3780} -\frac{71 a_{2}}{22680} + \frac{277 i m q}{22680} \right)z^5 +  \left( -\frac{31 i}{3780} - \frac{5 i a_{2}}{4536} - \frac{7 m q}{1620} \right)z^6 + \left( \frac{17}{5670} + \frac{79 a_{2}}{272160}\right.\right.\\
			\nonumber&\left.\left. - \frac{61 i m q}{54432} \right) z^7 +  \left( \frac{41 i}{54432} + \frac{7 i a_{2}}{116640} + \frac{47 m q}{204120} \right) z^8+ \left( -\frac{1579}{10692000} - \frac{37 a_{2}}{3592512} + \frac{703 i m q}{17962560} \right)z^9\right.\\
			\nonumber&\left.+\left(-\frac{i}{42000}-\frac{i a_2}{665280}-\frac{17mq}{2993760}\right)z^{10}+\left(\frac{29767}{9081072000}+\frac{1073a_2}{5604318720}-\frac{4027imq}{5604318720}\right)z^{11}\right.\\
			\nonumber&\left.+\left(\frac{14309i}{36324288000}+\frac{11ia_2}{509483520}+\frac{113mq}{1401079680}\right)z^{12}\right] +\eta^2\left[\left(\frac{1}{30}+\frac{a_2}{180}+\frac{a_2^2}{810}+\frac{a_3}{432}+\left(\frac{im}{40}\right.\right.\right.\\
			\nonumber&\left.\left.\left.-\frac{ima_2}{270}\right)q+\left(\frac{1}{60}-\frac{m^2}{240}\right)q^2\right)z^2+\left(-\frac{i}{60}+\frac{ia_2}{540}+\frac{ia_2^2}{540}+\frac{ia_3}{288}+\left(-\frac{m}{30}+\frac{ma_2}{270}\right)q+\left(\frac{i}{90}\right.\right.\right.\\
			\nonumber&\left.\left.\left.-\frac{im^2}{120}\right)q^2\right)z^3+\left(\frac{7937}{55125}-\frac{107lnz}{6300}-\frac{16609a_2}{793800}+\frac{a_2 lnz}{630}+\frac{4447a_2^2}{2976750}-\frac{8a_2^2 lnz}{14175}+\frac{823a_3}{317520}\right.\right.\\
			\nonumber&\left.\left.-\frac{a_3 lnz}{756}+\left(-\frac{53im}{9072}+\frac{5im a_2}{2268}\right)q+\left(-\frac{15247}{2646000}+\frac{lnz}{1575}+\frac{331319m^2}{47628000}-\frac{m^2 lnz}{14175}\right)q^2\right)z^4+\left(\frac{4673i}{55125}\right.\right.\\
			\nonumber&\left.\left.-\frac{107ilnz}{9450}-\frac{1931ia_2}{132300}+\frac{ia_2lnz}{945}+\frac{52481ia_2^2}{35721000}-\frac{16ia_2^2 lnz}{42525}+\frac{10189ia_3}{3810240}-\frac{ia_3lnz}{1134}+\left(-\frac{13m}{2835}\right.\right.\right.\\
			\nonumber&\left.\left.\left.-\frac{22ma_2}{25515}\right)q+\left(-\frac{17089i}{7938000}+\frac{2ilnz}{4725}+\frac{470803im^2}{142884000}-\frac{2im^2lnz}{42525}\right)q^2\right)z^5+\left(-\frac{1665983}{55566000}+\frac{1177lnz}{264600}\right.\right.\\
			\nonumber&\left.\left.+\frac{44357a_2}{7408800}-\frac{11a_2lnz}{26460}-\frac{12179a_2^2}{18522000}+\frac{44a_2^2lnz}{297675}-\frac{64997a_3}{53343360}+\frac{11a_3lnz}{31752}+\left(-\frac{1777im}{544320}\right.\right.\right.\\
			\nonumber&\left.\left.\left.-\frac{103ima_2}{408240}\right)q+\left(\frac{12017}{18522000}-\frac{11lnz}{66150}-\frac{4314983m^2}{4000752000}+\frac{11m^2lnz}{595350}\right)q^2\right)z^6+\left(-\frac{412849i}{55566000}\right.\right.\\
			\nonumber&\left.\left.+\frac{107ilnz}{88200}+\frac{112879ia_2}{66679200}-\frac{ia_2lnz}{8820}-\frac{580819ia_2^2}{3000564000}+\frac{4ia_2^2lnz}{99225}-\frac{116111ia_3}{320060160}+\frac{ia_3lnz}{10584}+\left(\frac{157m}{136080}\right.\right.\right.\\
			\nonumber&\left.\left.\left.+\frac{71ma_2}{1224720}\right)q+\left(\frac{26009i}{166698000}-\frac{ilnz}{22050}-\frac{802943im^2}{3000564000}+\frac{im^2lnz}{198450}\right)q^2\right)z^7+\left(\frac{47615027}{33006204000}\right.\right.\\
			\nonumber&\left.\left.-\frac{2461lnz}{9525600}-\frac{24542281a_2}{66012408000}+\frac{23a_2lnz}{952560}+\frac{402373a_2^2}{9282994875}-\frac{23a_2^2lnz}{2679075}+\frac{324463a_3}{3960744480}-\frac{23a_3lnz}{1143072}\right.\right.\\
			\nonumber&\left.\left.+\left(\frac{769619im}{2694384000}+\frac{887ima_2}{80831520}\right)q+\left(-\frac{4120093}{132024816000}+\frac{23lnz}{2381400}+\frac{11535901m^2}{216040608000}\right.\right.\right.\\
			\nonumber&\left.\left.\left.-\frac{23m^2lnz}{21432600}\right)q^2\right)z^8+\left(\frac{15116219i}{66012408000}-\frac{107ilnz}{2381400}-\frac{547717ia_2}{8251551000}+\frac{ia_2lnz}{238140}+\frac{4655411ia_2^2}{594111672000}\right.\right.\\
			\nonumber&\left.\left.-\frac{4ia_2^2lnz}{2679075}+\frac{944639ia_3}{63371911680}-\frac{ia_3lnz}{285768}+\left(-\frac{4661m}{84199500}-\frac{71ma_2}{40415760}\right)q+\left(-\frac{695159i}{132024816000}\right.\right.\right.\\
			\nonumber&\left.\left.\left.+\frac{ilnz}{595350}+\frac{21143443im^2}{2376446688000}-\frac{im^2lnz}{5358150}\right)q^2\right)z^9\right]+\eta^3\left[\left(-\frac{a_2}{90}+\frac{a_2^2}{3240}+\frac{a_3}{270}+\left(-\frac{im}{180}-\frac{ima_2}{1080}\right.\right.\right.\\
			\nonumber&\left.\left.\left.-\frac{ima_2^2}{972}-\frac{5ima_3}{2592}\right)q+\left(-\frac{1}{60}+\frac{m^2}{240}+\frac{43a_2}{15120}-\frac{\mathcal{H}_{(2m)}a_2}{1080}-\frac{43m^2a_2}{30240}\right)q^2\right)z+\left(\frac{i}{120}+\frac{ia_2}{720}\right.\right.\\
			\nonumber&\left.\left.+\frac{7ia_2^2}{19440}+\frac{7ia_2^3}{7290}+\frac{5ia_3}{5184}+\frac{13ia_2a_3}{3888}+\left(\frac{2m}{135}+\frac{7ma_2}{1296}+\frac{19ma_2^2}{14580}+\frac{19ma_3}{7776}\right)q+\left(-\frac{i}{360}+\frac{19im^2}{1440}\right.\right.\right.\\
			\nonumber&\left.\left.\left.-\frac{ia_2}{3780}-\frac{25im^2a_2}{54432}\right)q^2\right)z^2+\left(-\frac{10933}{49000}+\frac{107lnz}{3150}+\frac{31679a_2}{661500}-\frac{227a_2lnz}{37800}-\frac{25019a_2^2}{2976750}+\frac{79a_2^2lnz}{56700}\right.\right.\\
			\nonumber&\left.\left.+\frac{11167a_2^3}{17860500}-\frac{4a_2^3lnz}{42525}-\frac{12107a_3}{1270080}+\frac{a_3lnz}{378}+\frac{757a_2a_3}{476280}-\frac{a_2a_3lnz}{4536}+\left(-\frac{578569im}{7938000}+\frac{107imlnz}{12600}\right.\right.\right.\\
			\nonumber&\left.\left.\left.+\frac{5933ima_2}{529200}-\frac{ima_2lnz}{1260}-\frac{2381ima_2^2}{2976750}+\frac{4ima_2^2lnz}{14175}-\frac{289ima_3}{317520}+\frac{ima_3lnz}{1512}\right)q+\left(-\frac{9673}{1323000}\right.\right.\right.\\
			\nonumber&\left.\left.\left.-\frac{2lnz}{1575}-\frac{212879m^2}{23814000}+\frac{2m^2lnz}{14175}-\frac{10321a_2}{7938000}+\frac{\mathcal{H}_{(2m)}a_2}{7560}+\frac{a_2lnz}{9450}+\frac{175859m^2a_2}{285768000}\right.\right.\right.\\
			\nonumber&\left.\left.\left.-\frac{m^2a_2lnz}{85050}\right)q^2\right)z^3+\left(-\frac{52229i}{1323000}+\frac{107ilnz}{12600}+\frac{89149ia_2}{3969000}-\frac{137ia_2lnz}{37800}-\frac{79291ia_2^2}{20412000}+\frac{7ia_2^2lnz}{8100}\right.\right.\\
			\nonumber&\left.\left.-\frac{23ia_2^3}{425250}+\frac{ia_2^3lnz}{6075}-\frac{66557ia_3}{38102400}+\frac{19ia_3lnz}{15120}-\frac{1069ia_2a_3}{1905120}+\frac{ia_2a_3lnz}{1512}+\left(\frac{370721m}{3969000}-\frac{323mlnz}{37800}\right.\right.\right.\\
			\nonumber&\left.\left.\left.-\frac{15241ma_2}{1020600}+\frac{ma_2lnz}{540}+\frac{5563ma_2^2}{5953500}-\frac{11ma_2^2lnz}{85050}+\frac{16307ma_3}{9525600}-\frac{23ma_3lnz}{45360}\right)q+\left(-\frac{222013i}{31752000}\right.\right.\right.\\
			\nonumber&\left.\left.\left.-\frac{ilnz}{3150}-\frac{10211ia_2}{12348000}-\frac{11ia_2lnz}{132300}+\frac{691i\mathcal{H}_	{(2m)}a_2}{7938000}+\frac{i\mathcal{H}_{(2m)}a_2lnz}{37800}-\frac{32339im^2}{35721000}+\frac{im^2lnz}{28350}\right.\right.\right.\\
			\nonumber&\left.\left.\left.+\frac{63269im^2a_2}{222264000}+\frac{29im^2a_2lnz}{2381400}\right)q^2\right)z^4+\left(-\frac{4289239}{83349000}+\frac{4387 lnz}{793800}-\frac{51484a_2}{31255875}\right.\right.\\
			\nonumber&\left.\left.+\frac{5137a_2lnz}{4762800}+\frac{461869a_2^2}{1200225600}-\frac{253a_2^2lnz}{1428840}+\frac{23369a_2^3}{150028200}-\frac{64a_2^3lnz}{535815}-\frac{206063a_3}{88905600}+\frac{2a_3lnz}{59535}\right.\right.\\
			\nonumber&\left.\left.+\frac{4463a_2a_3}{6001128}-\frac{265a_2a_3lnz}{571536}+\left(\frac{23890883im}{500094000}-\frac{20819imlnz}{4762800}-\frac{10234787ima_2}{1200225600}+\frac{529ima_2lnz}{476280}\right.\right.\right.\\
			\nonumber&\left.\left.\left.+\frac{4318847ima_2^2}{9001692000}-\frac{113ima_2^2lnz}{2679075}+\frac{4872647ima_3}{4800902400}-\frac{671ima_3lnz}{2857680}\right)q+\left(\frac{1299007}{333396000}-\frac{41lnz}{198450}\right.\right.\right.\\
			\nonumber&\left.\left.\left.+\frac{1516189a_2}{4000752000}+\frac{79a_2lnz}{1190700}-\frac{12979\mathcal{H}_{(2m)}a_2}{285768000}-\frac{\mathcal{H}_{(2m)}a_2lnz}{56700}-\frac{5266951m^2}{4000752000}+\frac{41m^2lnz}{1786050}\right.\right.\right.\\
			\nonumber&\left.\left.\left.-\frac{522101m^2a_2}{4800902400}-\frac{m^2a_2lnz}{107163}\right)q^2\right)z^5+\left(-\frac{11209211i}{333396000}+\frac{3317ilnz}{793800}+\frac{541409ia_2}{266716800}+\frac{163ia_2lnz}{952560}\right.\right.\\
			\nonumber&\left.\left.-\frac{3477461ia_2^2}{12002256000}+\frac{23ia_2^2lnz}{7144200}+\frac{46337ia_2^3}{500094000}-\frac{263ia_2^3lnz}{5358150}-\frac{5185891ia_3}{3200601600}-\frac{323ia_3lnz}{1905120}\right.\right.\\
			\nonumber&\left.\left.+\frac{187877ia_2a_3}{480090240}-\frac{107ia_2a_3lnz}{571536}+\left(-\frac{2242607m}{142884000}+\frac{1003mlnz}{680400}+\frac{7533167ma_2}{2400451200}-\frac{197ma_2lnz}{476280}\right.\right.\right.\\
			\nonumber&\left.\left.\left.-\frac{199453ma_2^2}{1285956000}+\frac{13ma_2^2lnz}{1530900}-\frac{64577ma_3}{177811200}+\frac{419ma_3lnz}{5715360}\right)q+\left(\frac{1906043i}{1333584000}-\frac{31ilnz}{198450}\right.\right.\right.\\
			\nonumber&\left.\left.\left.+\frac{20243ia_2}{186701760}+\frac{19ia_2lnz}{666792}-\frac{13693i\mathcal{H}_{(2m)}a_2}{1000188000}-\frac{11i\mathcal{H}_{(2m)}a_2lnz}{1587600}-\frac{19515361im^2}{24004512000}+\frac{31im^2lnz}{1786050}\right.\right.\right.\\
			&\left.\left.\left.-\frac{14222783im^2a_2}{504094752000}-\frac{1181im^2a_2lnz}{300056400}\right)q^2\right)z^6\right].
		\end{align}
}}
\section{Expressions for \texorpdfstring{ $C_{ij}^q$}{} and \texorpdfstring{ $A_{ijq}$}{}}\label{ApenD}

In this part, we should first calculate the series expressions of $C_{nn}$, $C_{\overline{m}n}$ and $C_{\overline{mm}}$, which are given by
{\small{
		\begin{align}
			C_{n n}^{(0)} & =\frac{\Delta^0}{4 r^4 E}\left(E+\frac{d r}{d \tau}\right)^2, \\
			C_{n n}^{(1)}&
			=  \frac{L}{r^4} \frac{r^2-\Delta^0}{E} C_{n n}^{(0)}-\frac{L \Delta^0(E+d r / d \tau)}{2 E r^6}, \\
			C_{n n}^{(2)}&
			=  \frac{L}{r^4} \frac{r^2-\Delta^0}{E} C_{n n}^{(1)}+\mathcal{X}C_{n n}^{(0)}-\frac{\left(E+d r / d \tau\right)\left(E\cos^2\theta+d r / d \tau\right)\Delta^0}{2r^6 E}+\frac{L^2 \Delta^0}{4r^8 E},\\
			C_{\overline{m} n}^{(0)}&=\frac{\Delta^0}{2 \sqrt{2} r^3 E}\left(E+\frac{d r}{d \tau}\right)\left(\frac{{i} L}{r^2 \sin \theta}-\frac{d \theta}{d \tau}\right),\\
			C_{\overline{m} n}^{(1)}&=  \left(\frac{L}{r^4} \frac{r^2-\Delta^0}{E}+\frac{{i} \cos \theta}{r}\right) C_{\overline{m} n}^{(0)} -\frac{\Delta^0}{2 \sqrt{2} r^5 E}\left[{i} E \sin \theta\left(E+\frac{d r}{d \tau}\right)+L\left(\frac{{i} L}{r^2 \sin \theta}-\frac{d \theta}{d \tau}\right)\right],\\
			\nonumber
			C_{\overline{m} n}^{(2)}&=  \frac{L}{r^4} \frac{r^2-\Delta^0}{E} C_{\overline{m} n}^{(1)}+\left(\mathcal{X}-\frac{2}{r^2}-\frac{2\cos^2\theta}{r^2}\right) C_{\overline{m} n}^{(0)}-\frac{\cos\theta \left(r\cos\theta d\theta/d \tau-E\sin\theta\right)\Delta^0}{2\sqrt{2}r^6 E}\left(E+\frac{d r}{d \tau}\right)\\
			&+\frac{\Delta^0\left(rE-iL\cos\theta+r\cos^2\theta dr/d\tau\right)}{2\sqrt{2}r^6 E}\left(\frac{{i} L}{r^2 \sin \theta}-\frac{d \theta}{d \tau}\right)+\frac{iL\sin\theta\Delta^0}{2\sqrt{2}r^7},\\
			C_{\overline{m m}}^{(0)}&=\frac{\Delta^0}{2 r^2 E}\left(\frac{d\theta}{d\tau}-\frac{i L}{r^2 \sin \theta}\right)^2,\\
			C_{\overline{m m}}^{(1)}&=\left(\frac{L}{r^4} \frac{r^2-\Delta^0}{E}+\frac{2i \cos \theta}{r}\right) C_{\overline{m m}}^{(0)}+\frac{i\sin \theta \Delta^0}{r^4}\left(\frac{d \theta}{d \tau}-\frac{i L}{r^2 \sin \theta}\right),\\
			\nonumber C_{\overline{m m}}^{(2)}&=  \frac{L}{r^4} \frac{r^2-\Delta^0}{E} C_{\overline{m m}}^{(1)}+\left(\mathcal{X}-\frac{2}{r^2}-\frac{\cos^2\theta}{r^2}\right) C_{\overline{m m}}^{(0)}+\frac{i\cos\theta\left(L\cot\theta+2i r E\sin\theta\right)\Delta^0}{r^6 E}\\
			&\times\left(\frac{d \theta}{d \tau}-\frac{{i} L}{r^2 \sin \theta}\right)-\frac{E\sin^2\theta\Delta^0}{2r^6},
\end{align}}}
with

\begin{align}
	\mathcal{X}=\frac{1}{\Delta^0}+\frac{\sin^2\theta \Delta^0}{r^4}.
\end{align}

1. $\text{\large{\textbf{$A_{nn0}$}}}$

The coefficient $A_{nn0}^{i}(i=,0,1,2)$ of the source term can be expanded as
\begin{align}
	A_{nn0}=A_{nn0}^{0}+aA_{nn0}^{1}+a^{2}A_{nn0}^{2},
\end{align}
with
\begin{align}
	A_{nn0}^{0} & =-C_{nn}^{0}\cdot\frac{2r^{3}}{\sqrt{2\pi}\left(\Delta^{0}\right)^{2}}F_{6}^{0},\label{eq:a-nn0-0}\\
	A_{nn0}^{1} & =-C_{nn}^{1}\cdot\frac{2r^{3}}{\sqrt{2\pi}\left(\Delta^{0}\right)^{2}}F_{6}^{0}+C_{nn}^{0}\cdot\frac{2r^{2}}{\sqrt{2\pi}\left(\Delta^{0}\right)^{2}}\left(-rF_{6}^{1}+i\cos\theta F_{6}^{0}\right),\\
	A_{nn0}^{2} & =-C_{nn}^{2}\cdot\frac{2r^{3}}{\sqrt{2\pi}\left(\Delta^{0}\right)^{2}}F_{6}^{0}+C_{nn}^{1}\cdot\frac{2r^{2}}{\sqrt{2\pi}\left(\Delta^{0}\right)^{2}}\left(-rF_{6}^{1}+i\cos\theta F_{6}^{0}\right)\nonumber \\
	& \quad+C_{nn}^{0}\cdot\frac{2}{\sqrt{2\pi}\left(\Delta^{0}\right)^{2}}\left[-r^{3}F_{6}^{2}+ir^{2}\cos\theta F_{6}^{1}+\left(\frac{2r^{3}}{\Delta^{0}}-r\cos^2\theta\right)F_{6}^{0}\right], 
\end{align}
where
{\small{
		\begin{align}
			F_{6}^{0} & =rF_{2}^{0}\mathscr{L}_{1}^{0\dagger}\left[\mathscr{L}_{2}^{0\dagger}\left({}_{-2}Y_{\ell m}\right)\right],\\
			\nonumber	F_{6}^{1} & =\frac{i r F_{4}^{1}}{F_{5}^{0}}\mathscr{L}_{1}^{0\dagger}\left[\mathscr{L}_{2}^{0\dagger}\left[\cos\theta \left({}_{-2}Y_{\ell m}\right)\right]\right]+\frac{\omega r F_{4}^{0}}{F_{5}^{0}}\mathscr{L}_{1}^{0\dagger}\left[\sin\theta \left({}_{-2}Y_{\ell m}\right)\right]-\frac{i F_{4}^{0}\left(F_{5}^{0}+r F_{5}^{1}\right)}{\left(F_{5}^{0}\right)^{2}}\\
			&\times\mathscr{L}_{1}^{0\dagger}\left[\cos\theta\mathscr{L}_{2}^{0\dagger}\left({}_{-2}Y_{\ell m}\right)\right]+\frac{\omega r F_{4}^{0}}{F_{5}^{0}}\sin\theta\mathscr{L}_{2}^{0\dagger}\left({}_{-2}Y_{\ell m}\right),\\
			\nonumber	F_{6}^{2} & =\frac{r F_{4}^{2}}{F_{5}^{0}}\mathscr{L}_{1}^{0\dagger}\left[\mathscr{L}_{2}^{0\dagger}\left[\cos^2\theta ({}_{-2}Y_{\ell m})\right]\right]+\frac{i \omega r F_{4}^{1}}{F_{5}^{0}}\mathscr{L}_{1}^{0\dagger}\left[\sin\theta\cos\theta ({}_{-2}Y_{\ell m})\right]+\frac{F_{4}^{1} \left(F_{5}^{0}+r F_{5}^{1}\right)}{\left(F_{5}^{0}\right)^{2}}\\
			\nonumber	&\times\mathscr{L}_{1}^{0\dagger}\left[\cos \theta\mathscr{L}_{2}^{0\dagger}\left[\cos\theta ({}_{-2}Y_{\ell m})\right]\right]-\frac{i \omega F_{4}^{0}\left(F_{5}^{0}+r F_{5}^{1}\right)}{\left(F_{5}^{0}\right)^{2}}\mathscr{L}_{1}^{0\dagger}\left[ \sin\theta\cos\theta({}_{-2}Y_{\ell m})\right]\\
			\nonumber	&-\frac{F_{4}^{0}\left(F_{5}^{0}F_{5}^{1}+r\left(F_{5}^{1}\right)^{2}+rF_{5}^{0}F_{5}^{2}\right)}{\left(F_{5}^{0}\right)^{3}}\mathscr{L}_{1}^{0\dagger}\left[\cos^2\theta\mathscr{L}_{2}^{0\dagger}\left({}_{-2}Y_{\ell m}\right)\right]+\frac{i \omega r F_{4}^{1}}{F_{5}^{0}}\sin \theta\mathscr{L}_{2}^{0\dagger}\left[\cos\theta \left({}_{-2}Y_{\ell m}\right)\right]\\
			& \quad+\frac{\omega^2 r F_{4}^{0}}{F_{5}^{0}}\sin^2\theta \left({}_{-2}Y_{\ell m}\right)-\frac{i \omega F_{4}^{0}\left(F_{5}^{0}+r F_{5}^{1}\right)}{\left(F_{5}^{0}\right)^{2}}\sin\theta\cos \theta\mathscr{L}_{2}^{0\dagger}\left({}_{-2}Y_{\ell m}\right).
		\end{align}
}}

2. $\text{\large{\textbf{$A_{\overline{m}n0}$}}}$

The coefficient $A_{\overline{m}n0}^{i}(i=,0,1,2)$ of the source term can be expanded as
\begin{align}
	A_{\overline{m}n0}=A_{\overline{m}n0}^{0}+aA_{\overline{m}n0}^{1}+a^{2}A_{\overline{m}n0}^{2},
\end{align}
with
\begin{align}
	A_{\overline{m}n0}^{0} & =C_{\overline{m}n}^{0}\cdot\frac{r^{4}}{\sqrt{\pi}\Delta^{0}}F_{7}^{0},\label{eq:a-mn0-0}\\
	A_{\overline{m}n0}^{1} & =C_{\overline{m}n}^{1}\cdot\frac{r^{4}}{\sqrt{\pi}\Delta^{0}}F_{7}^{0}+C_{\overline{m}n}^{0}\cdot\frac{r^{4}}{\sqrt{\pi}\Delta^{0}}F_{7}^{1},\\
	A_{\overline{m}n0}^{2} & =C_{\overline{m}n}^{2}\cdot\frac{r^{4}}{\sqrt{\pi}\Delta^{0}}F_{7}^{0}+C_{\overline{m}n}^{1}\cdot\frac{r^{4}}{\sqrt{\pi}\Delta^{0}}F_{7}^{1}+C_{\overline{m}n}^{0}\cdot\frac{r^{2}}{\sqrt{\pi}\Delta^{0}}\left[r^{2}F_{7}^{2}+\left(2\cos^{2}\theta-\frac{r^{2}}{\Delta^{0}}\right)F_{7}^{0}\right],
\end{align}
where
{\small{
		\begin{align}
			F_{7}^{0} & =F_{4}^{0}\left(\frac{1}{rF_{4}^{0}}\right)_{\prime r}\mathscr{L}_{2}^{0\dagger}\left({}_{-2}Y_{\ell m}\right)-\mathscr{D}_{0}^{0}\left(\frac{1}{r}F_{9}^{0}\right),\\
			\nonumber	F_{7}^{1} & =\omega\sin\theta\cdot F_{4}^{0}\left(\frac{1}{rF_{4}^{0}}\right)_{\prime r}\left({}_{-2}Y_{\ell m}\right) +\left[i F_{4}^{1}\left(\frac{1}{rF_{4}^{0}}\right)_{\prime r}-F_{4}^{0}\left(\frac{i\left(3F_{4}^{0}+rF_{4}^{1}\right)}{r^{2}\left(F_{4}^{0}\right)^{2}}\right)_{\prime r}\right]\mathscr{L}_{2}^{0\dagger}\left[\cos\theta\left({}_{-2}Y_{\ell m}\right)\right]\\
			&-\frac{im}{r\Delta^{0}}F_{9}^{0}-\mathscr{D}_{0}^{0}\left(\frac{rF_{9}^{1}-3i\cos\theta F_{9}^{0}}{r^{2}}\right),\\
			\nonumber	F_{7}^{2} & =i\omega\cos\theta\sin\theta\left[ F_{4}^{1}\left(\frac{1}{rF_{4}^{0}}\right)_{\prime r}-F_{4}^{0}\left(\frac{i\left(3F_{4}^{0}+rF_{4}^{1}\right)}{r^{2}\left(F_{4}^{0}\right)^{2}}\right)_{\prime r}\right]\left({}_{-2}Y_{\ell m}\right)\\
			\nonumber	& +\left[F_{4}^{2}\left(\frac{1}{rF_{4}^{0}}\right)_{\prime r}-i F_{4}^{1}\left(\frac{i\left(3F_{4}^{0}+rF_{4}^{1}\right)}{r^{2}\left(F_{4}^{0}\right)^{2}}\right)_{\prime r}-F_{4}^{0}\left(\frac{\left[5\left(F_{4}^{0}\right)^{2}+r^{2}\left(F_{4}^{1}\right)^{2}+rF_{4}^{0}\left(3F_{4}^{1}+rF_{4}^{2}\right)\right]}{r^{3}\left(F_{4}^{0}\right)^{3}}\right)_{\prime r}\right]\\
			&\times\mathscr{L}_{2}^{0\dagger}\left[\cos^2 \theta\left({}_{-2}Y_{\ell m}\right)\right]-\mathscr{D}_{0}^{2}\frac{F_{9}^{0}}{r}-\frac{im}{\Delta^{0}}\cdot\frac{rF_{9}^{1}-3i\cos\theta F_{9}^{0}}{r^{2}}-\mathscr{D}_{0}^{0}\left(\frac{r^{2}F_{9}^{2}-3ir\cos\theta F_{9}^{1}-5\cos^{2}\theta F_{9}^{0}}{r^{3}}\right).
		\end{align}
}}

3. $\text{\large{\textbf{$A_{\overline{mm}}$}}}$

The coefficient $A_{\overline{mm}0}^{i}(i=,0,1,2)$ of the source term can be expanded as
\begin{align}
	A_{\overline{mm}0}=A_{\overline{mm}0}^{0}+aA_{\overline{mm}0}^{1}+a^{2}A_{\overline{mm}0}^{2},
\end{align}
with
\begin{align}
	A_{\overline{mm}0}^{0} & =-C_{\overline{mm}}^{0}\cdot\frac{r^{2}}{\sqrt{2\pi}}F_{8}^{0},\label{eq:a-mm0-0}\\
	A_{\overline{mm}0}^{1} & =-C_{\overline{mm}}^{1}\cdot\frac{r^{2}}{\sqrt{2\pi}}F_{8}^{0}+C_{\overline{mm}}^{0}\cdot\frac{r}{\sqrt{2\pi}}\left(-rF_{8}^{1}+4i\cos\theta F_{8}^{0}\right), \\
	A_{\overline{mm}0}^{2} & =-C_{\overline{mm}}^{2}\cdot\frac{r^{2}}{\sqrt{2\pi}}F_{8}^{0}+C_{\overline{mm}}^{1}\cdot\frac{r}{\sqrt{2\pi}}\left(-rF_{8}^{1}+4i\cos\theta F_{8}^{0}\right)\nonumber \\
	& \quad+C_{\overline{mm}}^{0}\cdot\frac{1}{\sqrt{2\pi}}\left(-r^{2}F_{8}^{2}+4ir\cos\theta F_{8}^{1}+7\cos^2\theta F_{8}^{0}\right),
\end{align}
where
{\small{
		\begin{align}
			F_{8}^{0} & =\left[-i\left(\frac{r^{2}\omega}{\Delta^{0}}\right)_{\prime r}-\frac{r^{4}\omega^{2}}{\left(\Delta^{0}\right)^{2}}-\frac{ir\left(1+rF_{1}^{0}\right)\omega}{\Delta^{0}}+\frac{F_{1}^{0}}{r}+\left(F_{1}^{0}\right)_{\prime r}\right]\left({}_{-2}Y_{\ell m}\right),\\
			\nonumber	F_{8}^{1} & =\Bigg\{i\left(\frac{m}{\Delta^{0}}\right)_{\prime r}+i\left(\cos\theta F_{1}^{1}\right)_{\prime r}+\frac{1}{r^{2}\left(\Delta^{0}\right)^{2}}\left\{ 2mr^{4}\omega+r\Delta^{0}\left[i\left(m+rmF_{1}^{0}\right)+r\left(1+r^{2}F_{1}^{1}\right)\omega\cos\theta\right]\right.\\
			& \quad\left.+i\cos\theta\left(\Delta^{0}\right)^{2}\left(F_{1}^{0}+rF_{1}^{1}\right)\right\}\Bigg\}\left({}_{-2}Y_{\ell m}\right) ,\\
			\nonumber	F_{8}^{2} & =\Bigg\{i\left(\frac{\omega\left(r^{2}-\Delta^{0}\right)}{\left(\Delta^{0}\right)^{2}}\right)_{\prime r}-\frac{\cos^2\theta\left[F_{1}^{0}+r\left(F_{1}^{1}+rF_{1}^{2}\right)\right]}{r^{3}}+\frac{2r^{4}\omega^{2}}{\left(\Delta^{0}\right)^{3}}+\frac{-m^{2}+ir\omega\left(1+rF_{1}^{0}+2ir\omega\right)}{\left(\Delta^{0}\right)^{2}}\\
			& \quad-\frac{1}{r^{2}\Delta^{0}}\left\{ ir\omega\left(1+rF_{1}^{0}\right)+\cos\theta\left[m+mr^{2}F_{1}^{1}+ir\omega\cos\theta\left(-1+r^{3}F_{1}^{2}\right)\right]\right\} +\left(\cos^2\theta F_{1}^{2}\right)_{\prime r}\Bigg\}\left({}_{-2}Y_{\ell m}\right).
		\end{align}
}}

4. $\text{\large{\textbf{$A_{\overline{m}n1}$}}}$

The coefficient $A_{\overline{m}n1}^{i}(i=,0,1,2)$ of the source term can be expanded as
\begin{align}
	A_{\overline{m}n1}=A_{\overline{m}n1}^{0}+aA_{\overline{m}n1}^{1}+a^{2}A_{\overline{m}n1}^{2},
\end{align}
with
\begin{align}
	A_{\overline{m}n1}^{0} & =C_{\overline{m}n}^{0}\cdot\frac{r^{3}}{\sqrt{\pi}\Delta^{0}}F_{9}^{0},\label{eq:a-mn1-0}\\
	A_{\overline{m}n1}^{1} & =C_{\overline{m}n}^{1}\cdot\frac{r^{3}}{\sqrt{\pi}\Delta^{0}}F_{9}^{0}+C_{\overline{m}n}^{0}\cdot\frac{r^{2}}{\sqrt{\pi}\Delta^{0}}\left(rF_{9}^{1}-3i\cos\theta F_{9}^{0}\right), \\
	A_{\overline{m}n1}^{2} & =C_{\overline{m}n}^{2}\cdot\frac{r^{3}}{\sqrt{\pi}\Delta^{0}}F_{9}^{0}+C_{\overline{m}n}^{1}\cdot\frac{r^{2}}{\sqrt{\pi}\Delta^{0}}\left(rF_{9}^{1}-3i\cos\theta F_{9}^{0}\right)\nonumber \\
	& \quad+C_{\overline{m}n}^{0}\cdot\frac{1}{\sqrt{\pi}\Delta^{0}}\left[r^{3}F_{9}^{2}-3ir^{2}\cos\theta F_{9}^{1}-\left(\frac{r^{3}}{\Delta^{0}}+3r\cos^2\theta\right)F_{9}^{0}\right], 
\end{align}
where
\begin{align}
	F_{9}^{0} & =\left(1+F_{2}^{0}\right)\mathscr{L}_{2}^{0\dagger}\left({}_{-2}Y_{\ell m}\right),\\
	F_{9}^{1} & =\left[-\frac{i\sin\theta F_{4}^{1}}{F_{5}^{0}}+\frac{3i\sin\theta}{r}+i\cos\theta F_{2}^{1}\mathscr{L}_{2}^{0\dagger}+\left(1+F_{2}^{0}\right)\omega\sin\theta\right]\left({}_{-2}Y_{\ell m}\right),\\
	F_{9}^{2} & =\left[\cos^2\theta F_{2}^{2}\mathscr{L}_{2}^{0\dagger}+i\omega\cos\theta\sin\theta F_{2}^{1}+\frac{\cos\theta\sin\theta}{r^{2}}-\frac{\left(2F_{4}^{2}F_{5}^{0}+F_{4}^{1}F_{5}^{1}\right)\cos\theta\sin\theta}{\left(F_{5}^{0}\right)^{2}}\right]\left({}_{-2}Y_{\ell m}\right).
\end{align}

5. $\text{\large{\textbf{$A_{\overline{mm}1}$}}}$

The coefficient $A_{\overline{mm}1}^{i}(i=,0,1,2)$ of the source term can be expanded as
\begin{align}
	A_{\overline{mm}1}=A_{\overline{mm}1}^{0}+aA_{\overline{mm}1}^{1}+a^{2}A_{\overline{mm}1}^{2},
\end{align}
with
\begin{align}
	A_{\overline{mm}1}^{0} & =C_{\overline{mm}}^{0}\cdot\frac{r^{2}}{\sqrt{2\pi}}F_{10}^{0},\label{eq:a-mm1-0}\\
	A_{\overline{mm}1}^{1} & =C_{\overline{mm}}^{1}\cdot\frac{r^{2}}{\sqrt{2\pi}}F_{10}^{0}+C_{\overline{mm}}^{0}\cdot\frac{r}{\sqrt{2\pi}}\left(rF_{10}^{1}-4iF_{10}^{0}\cos\theta\right) \\
	A_{\overline{mm}1}^{2} & =C_{\overline{mm}}^{2}\cdot\frac{r^{2}}{\sqrt{2\pi}}F_{10}^{0}+C_{\overline{mm}}^{1}\cdot\frac{r}{\sqrt{2\pi}}\left(rF_{10}^{1}-4iF_{10}^{0}\cos\theta\right)\nonumber \\
	& \quad+C_{\overline{mm}}^{0}\cdot\frac{1}{\sqrt{2\pi}}\left(r^{2}F_{10}^{2}-4irF_{10}^{1}\cos\theta-7F_{10}^{0}\cos^{2}\theta\right),
\end{align}
where
\begin{align}
	F_{10}^{0} & =\left(F_{1}^{0}+\frac{1}{r}-i\frac{2\omega}{\Delta^{0}}r^{2}\right)\left({}_{-2}Y_{\ell m}\right),\\
	F_{10}^{1} & =i\left(F_{1}^{1}\cos\theta+\frac{\cos\theta}{r^{2}}+\frac{2m}{\Delta^{0}}\right)\left({}_{-2}Y_{\ell m}\right),\\
	F_{10}^{2} & =\left[\left(F_{1}^{2}-\frac{1}{r^{3}}\right)\cos^{2}\theta+i\frac{2\omega\left(r^{2}-\Delta^{0}\right)}{\left(\Delta^{0}\right)^{2}}\right]\left({}_{-2}Y_{\ell m}\right).
\end{align}

6. $\text{\large{\textbf{$A_{\overline{mm}2}$}}}$

The coefficient $A_{\overline{mm}2}^{i}(i=,0,1,2)$ of the source term can be expanded as
\begin{align}
	A_{\overline{mm}2}=A_{\overline{mm}2}^{0}+aA_{\overline{mm}2}^{1}+a^{2}A_{\overline{mm}2}^{2},
\end{align}
with
\begin{align}
	A_{\overline{mm}2}^{0} & =-C_{\overline{mm}}^{0}\cdot\frac{r^{2}}{\sqrt{2\pi}}\left({}_{-2}Y_{\ell m}\right),\label{eq:a-mm2-0}\\
	A_{\overline{mm}2}^{1} & =-C_{\overline{mm}}^{1}\cdot\frac{r^{2}}{\sqrt{2\pi}}\left({}_{-2}Y_{\ell m}\right)+C_{\overline{mm}}^{0}\cdot\frac{4ir\cos\theta}{\sqrt{2\pi}}\left({}_{-2}Y_{\ell m}\right), \\
	A_{\overline{mm}2}^{2} & =-C_{\overline{mm}}^{2}\cdot\frac{r^{2}}{\sqrt{2\pi}}\left({}_{-2}Y_{\ell m}\right)+C_{\overline{mm}}^{1}\cdot\frac{4ir\cos\theta}{\sqrt{2\pi}}\left({}_{-2}Y_{\ell m}\right)+C_{\overline{mm}}^{0}\cdot\frac{7\cos^{2}\theta}{\sqrt{2\pi}}\left({}_{-2}Y_{\ell m}\right).
\end{align}
%\end{appendices}

\end{document}